# Ethylenediamine Addition Improves Performance and Suppresses Phase Instabilities in Mixed-Halide Perovskites

*Margherita Taddei,[1] Joel A. Smith,[2] Benjamin M. Gallant,[2] Suer Zhou,[2] Robert J. E. Westbrook,[1] Yangwei Shi,[1] Jian Wang,[1] James N. Drysdale,[2] Declan P. McCarthy,[3] Stephen Barlow,[3] Seth R. Marder,[3,4] Henry J. Snaith[2] and David S. Ginger[1]\**

AUTHOR ADDRESS**:** Department of Chemistry, University of Washington, Seattle, WA, USA

AUTHOR INFORMATION

1 Department of Chemistry, University of Washington, Seattle WA, USA

2 Department of Physics, University of Oxford, Oxford, UK

3 Renewable and Sustainable Energy Institute, University of Colorado Boulder, Boulder CO, USA

4 Department of Chemical and Biological Engineering and Department of Chemistry, University of Colorado Boulder, Boulder CO, USA

**Corresponding Author**

\* David S. Ginger, Department of Chemistry, University of Washington, Seattle WA, USA

**ABSTRACT:**

We show that adding ethylenediamine (EDA) to perovskite precursor solution improves the photovoltaic device performance and material stability of high-bromide-content, methylammonium-free, formamidinium cesium lead halide perovskites $FA_{1-x}Cs_xPb(I_{1-y}Br_y)_3$ which are currently of interest for perovskite-on-Si tandem solar cells. Using spectroscopy and hyperspectral microscopy, we show that the additive improves film homogeneity and suppresses the phase instability that is ubiquitous in high-Br perovskite formulations, producing films that remain stable for over 100 days in ambient conditions. With the addition of 1 mol% EDA we demonstrate 1.69 eV-gap perovskite single-junction p-i-n devices with a $V_{OC}$ of 1.22 V, and a champion maximum power point tracked power conversion efficiency of 18.8%, comparable to the best reported methylammonium-free perovskites. Using nuclear magnetic resonance (NMR) spectroscopy and X-ray diffraction techniques, we show that EDA reacts with $FA^+$ in solution, rapidly and quantitatively forming imidazolinium cations. It is the presence of imidazolinium during crystallization which drives the improved perovskite thin-film properties.

**TOC**

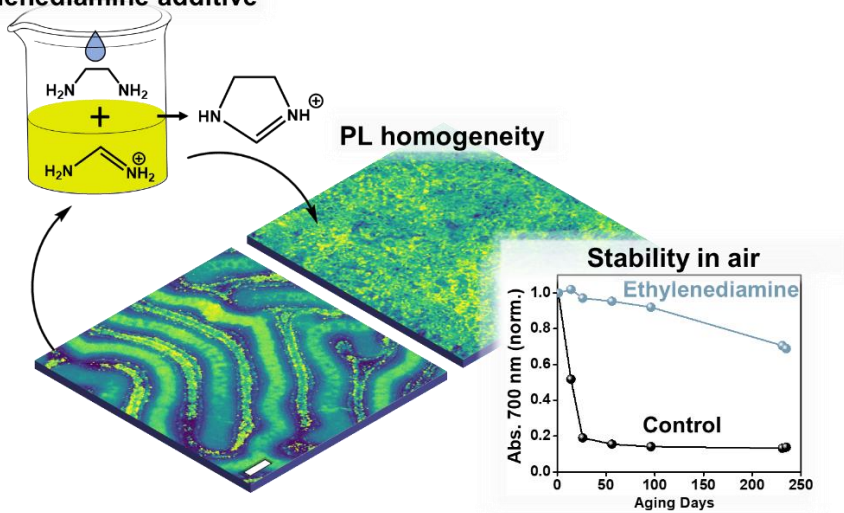

Metal halide perovskites (MHPs) combine excellent semiconductor properties with the potential for scalable, low-cost manufacturing. As a result, these materials have attracted attention in optoelectronic applications ranging from photovoltaics[1], light-emitting diodes[2], and detectors[3] to elements of non-linear optics and sources of quantum light.[4] MHPs have the generic formula $ABX_3$ – where A (= $Cs^+$, methylammonium or $MA^+$, formamidinium or $FA^+$) and B (= $Pb^{2+}$, $Sn^{2+}$) are cations and X (= $I^-$, $Br^-$, $Cl^-$) is an anion and offer broad band gap tunability from <1.25 eV in e.g. $(FASnI_3)_{0.6}(MAPbI_3)_{0.4}$ to >2.25 eV in e.g. $CsPbBr_3$.[5,6] In photovoltaics, there is particular interest in overcoming the single-junction, detailed-balance efficiency limit using multijunction solar cells.[7,8]

The ideal bandgap for the perovskite active layer in Si-perovskite tandem solar cells is roughly 1.7 eV.[9] In order to achieve this bandgap for typical multi-cation compositions, iodide is partially replaced with bromide at the halide ($X^-$) site, generally using over 23% $Br^-$.[10] However, perovskites with high bromide/iodide ratios (>20%) usually undergo "halide segregation", by which iodide-rich narrower bandgap regions are formed during illumination, compromising the performance and stability of the solar cell.[11,12] In recent years, due to thermal instabilities of methylammonium perovskites, compositions containing only formamidinium and cesium ($FA^+/Cs^+$) cations have been increasingly favored by the community.[13,14] However, these $FA^+/Cs^+$ compositions suffer from phase instabilities over time: the desired cubic corner-sharing (α) "black" phases separate to pure $CsPbI_3$ and $FAPbI_3$, which are 1D photoinactive, non-perovskite "yellow" phases in which the $BX_6$ octahedra are edge-sharing and face-sharing respectively.[15] Thus, it is necessary to find methods to limit both halide segregation and phase instability to achieve stable wide-gap compositions.

Diamines have been previously incorporated into metal halide perovskites as additives in the precursor solution.[16–19] In particular, the addition of ethylenediamine (EDA) and ethylene diammonium iodide (EDAI$_2$) has been shown to lead to the formation of "hollow" materials in certain ABI$_3$ (A = MA$^+$, FA$^+$; B = Pb$^{2+}$, Sn$^{2+}$) perovskites, whereby diammonium di-cations are incorporated into the perovskite structure, along with B$^{2+}$ and X$^-$ vacancies to enable steric accommodation of the EDA di-cations and balancing of the overall charge respectively.[20,21] In addition, the use of EDA via post-treatment of the surface has been reported to afford increased efficiency and stability in Pb-Sn perovskites.[22] Likewise, dimethylammonium is another bulky cation, which is too large to incorporate into many halide perovskite compositions according to the Goldschmidt tolerance factor considerations, but in some cases has been shown to increase both the bandgap, open-circuit voltage ($V_{OC}$), and overall performance of perovskite solar cells, despite negatively impacting the compositional heterogeneity of those films.[23–25] Motivated by these observations, we hypothesized that the addition of EDA might be beneficial for formulations close to the 1.7 eV band gap ideal for silicon-perovskite tandems. We anticipated that the addition of EDA might allow us to access a more stable compositional space, thus allowing the use of lower bromide concentrations to achieve the 1.7 eV band gap ideal for silicon-perovskite tandems.

Here, we demonstrate that incorporating EDA into a >1.68 eV bandgap FA$_{0.83}$Cs$_{0.17}$Pb(I$_{0.75}$Br$_{0.25}$)$_3$ perovskite yields a significant enhancement in both chemical homogeneity and phase stability, as well as an enhancement of solar cell device performance. We show p-i-n devices with a maximum power point tracked power conversion efficiency (PCE) of 18.8% and open circuit voltage ($V_{OC}$) of 1.22 V with 1 mol% EDA (EDA-1) addition which is a record for this bandgap in a MA-free composition. We investigate the mechanism for these improvements using nuclear magnetic resonance (NMR) spectroscopy and X-ray diffraction

(XRD) techniques and find that EDA reacts with $FA^+$ in solution under the conditions employed here.

First, we investigate the effects of adding EDA to the growth solution used to deposit films of the composition $FA_{0.83}Cs_{0.17}Pb(I_{0.75}Br_{0.25})_3$, which has a bandgap close to the ideal top cell for a Si-tandem ($E_g$= 1.695 eV, extracted from UV-Vis of **Figure 1a**). **Figure 1a** shows that addition of 1 mol% EDA (EDA-1) as an additive in the precursor solution used to spin-coat the films causes a very slight blueshift in the absorption ($E_g$=1.702 eV) compared to those of the same composition without EDA (EDA-0), along with an increase in measured photoluminescence intensity. PL measurements in an integrating sphere at an intensity of 60 mW/cm$^2$ with a 532 nm laser indicate that the EDA-1 film has a photoluminescence quantum efficiency (PLQE) of 9.7% and the emission maximum ($\lambda_{max}$) is located at 717 nm, whereas the EDA-0 sample has a PLQE of 6.5%, with its maximum at 722 nm (**Figure S1**). **Figure 1a** also shows that the EDA-0 film has a broader PL peak, with a clear shoulder at 744 nm. We observe this shoulder growing rapidly in EDA-0 samples, even during a brief PL measurement (~timescale here 0.1s), consistent with reports of the facile segregation into I-rich and Br-rich phases for such high-bromide composition samples.[26–28]

**Figure 1b** shows the PL lifetime decays of the EDA-0 and EDA-1 films taken using 405 nm excitation with a pulse energy/area of ~4 nJ cm$^{-2}$ ($N_0$= 1.8 × 10$^{14}$ cm$^{-3}$). Under these conditions the differences in PL dynamics between the reference and EDA-1 films is clear, with the EDA-1 film showing significantly more counts, and a longer PL lifetime. We fit the PL decays using a stretched exponential function, as we have reported previously [29,30]:

$$I(t) = A_1 e^{-\left(\frac{t}{\tau_c}\right)^\beta} \tag{1}$$

We use the stretched exponential to account phenomenologically for distributions of PL lifetimes that carriers excited at different lateral and vertical positions in the film may experience due to sample heterogeneity. The stretched exponential fits for the reference and EDA-1 films give $\tau_c$ and beta values of $\tau_c$ =0.6 ns, and $\beta$ =0.17 for the reference film, and $\tau_c$ =301.5 ns and $\beta$ =0.49 for the EDA-1 films (see **Table S1**). The stretching exponent $\beta$ is indicative of the width of the lifetime distribution in an ensemble sample, with a $\beta$=1 recovering a perfectly homogeneous distribution of single-exponential emitters. Thus, the higher $\beta$ value of the EDA-1 film (0.49) compared to that of the EDA-0 film (0.17) indicates a narrower distribution of emissive states, with the longer average lifetime $\tau_{stretch}$ calculated via:

$$\tau_{stretch} = \frac{\tau_c}{\beta} \Gamma\left(\frac{1}{\beta}\right) \tag{2}$$

of 627 ns for the EDA-1 film compared to 411 ns for the EDA-0 film indicating a lower distribution of non-radiative recombination pathways in the EDA-1 film.

Next, we monitor the phase stability of unencapsulated films over time whilst storing them in ambient air, in the dark, at room temperature (ISOS D-1 conditions).[31] **Figure 1c** shows the EDA-0 films rapidly (~days to weeks) revert to non-corner sharing phases in ambient conditions, as previously reported,[15] whereas the EDA-1 film is still predominantly in the desired photo-active black phase after 235 days. To quantify this stability, we use the UV-Vis absorbance of the film at 700 nm as a proxy for the remaining pseudo-cubic perovskite. **Figure 1d** plots the resulting absorbance at 700 nm for each film (full UV-Vis spectra are shown in **Figure S2**). As is typical, the EDA-0 film shows a rapid, steep drop in optical density to less than 20% of the initial value after 25 days. In contrast, the EDA-1 maintains 96% of its initial absorbance after 100 days in ambient. To characterize the phases formed during aging, we acquired XRD patterns of fresh films and after 100 days of aging in the same ambient conditions. In **Figure 1e** (full XRD patterns shown in **Figure S3**) the EDA-0 film shows the formation of several decomposition products including

FAPbX$_3$ polytype phases, orthorhombic δ-CsPbX$_3$,[32] and peaks consistent with a proposed CsPb$_2$I$_4$Br phase (**Figure S4**).[33] In contrast, EDA-1 shows no additional phases detectable by XRD after 100 days of aging. The stability was also tested under a more accelerated condition with ambient light and air flow where we note that stability was further enhanced with 10 mol% EDA (EDA-10), shown in **Figure S5**.

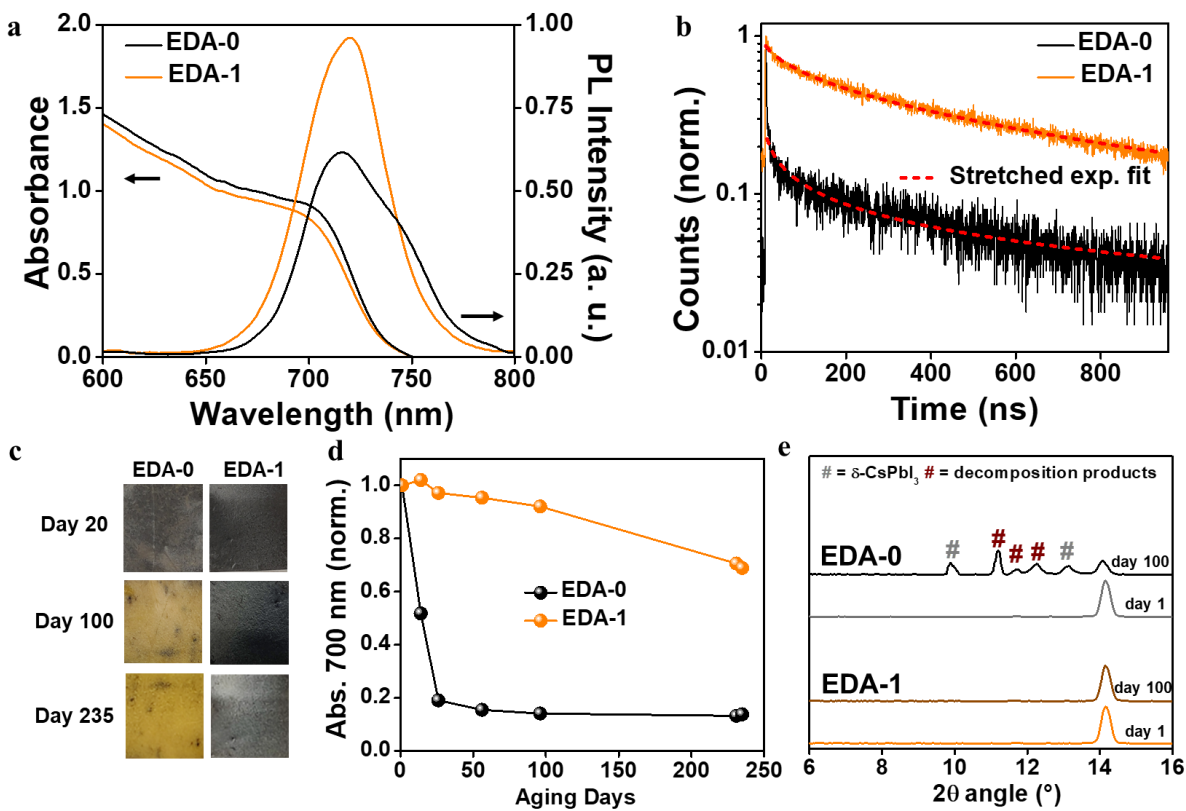

*Figure 1*: **Material characterization and phase stability**. *(a) UV-Vis spectra and normalized PL emission from films of FA$_{0.83}$Cs$_{0.17}$Pb(I$_{0.75}$Br$_{0.25}$)$_3$ without EDA (EDA-0, black) and with 1% EDA (EDA-1, orange). The PL spectra are obtained using an integrating sphere coupled with a 532 nm laser at 1 Sun fluence. (b) Time resolved photoluminescence measurements using a 405 nm laser at 1 MHz pulse frequency. The fitting (dashed red line) was done using a stretched exponential decay. (c) Visual degradation of EDA-0 and EDA-1 exposed to air for 20, 100 and 235 days.*

*Samples were left unencapsulated in ambient air in the dark at an average room temperature of 19.5°C and relative humidity of 31%. (d) Absorbance from UV-Vis of EDA-0 (black) and EDA-1 (orange) at 700 nm during ambient aging. (e) XRD patterns in the 2θ = 6°-16° region of the fresh EDA-0 (grey) and EDA-1(orange) and after 100 days of ambient aging (EDA-0: black, EDA-1: brown).*

Given the promising increases in stability, PLQY, and PL lifetime, we fabricated p-i-n solar cells using EDA as an additive, with the device architecture glass/ITO/Me-4PACz/np-Al$_2$O$_3$/perovskite/PCBM/BCP/Au, where Me-4PACz is a carbazole-based self-assembled monolayer (SAM) hole extraction layer ([4-(3,6-dimethyl-9H-carbazol-9-yl)butyl]phosphonic acid), np-Al$_2$O$_3$ is a dilute nanoparticle alumina layer employed to improve the perovskite solution wetting, PCBM ([6,6]-phenyl-C61-butyric acid methyl ester) is the electron extraction layer, and BCP (bathocuproine) rectifies the fullerene/metal junction.[34] Employing perovskite precursor solutions with a range of EDA additions from 0 - 3 mol% had a marked effect on the device performance measured under standard conditions (full details given in the Methods). **Figure 2a** shows stabilized maximum power point (MPP) efficiencies of the resulting devices (0.25 cm$^2$ active area), with the MPP shown for more representative performance comparison.[35] The full current-voltage (*JV*) performance metrics – power conversion efficiency (PCE), short-circuit current density ($J_{SC}$), fill factor (FF) and $V_{OC}$ – are shown in **Figure S6.** The $V_{OC}$ increases steadily with EDA addition over the concentration series, which we propose could be due to a combination of passivation and/or blueshift effects with the additive. Both $J_{SC}$ and FF reach a maximum at 1% EDA, with a particularly severe decline in $J_{SC}$ above this optimum value.

**Figure 2b** shows the *JV* results from the best EDA-1 device with a maximum PCE of 19.07%, $V_{OC}$ of 1.22V, $J_{SC}$ of 19.1 mA/cm$^2$, and FF of 82%, with little observable hysteresis. We verified the measured $J_{SC}$ using external quantum efficiency (EQE) measurements, with the integrated current over the spectral range in perfect agreement with the *JV*-determined $J_{SC}$ (**Figure S7**). The results for the EDA-1 device were compared to the literature on MA-free wide-gap compositions (**Table S2**).

Our champion 19.07% PCE for a 1.685 eV bandgap composition (**Figure 2c**) compares favorably to other reports, in particular the high $V_{OC}$ of 1.22 V (**Figure 2d**). Additionally, in **Figure S8** we show the performance for larger 1 cm$^2$ devices fabricated in the same batches and on the same device substrates as the small area cells. Notably, EDA-1 devices show significantly higher average and champion efficiencies compared to without the additive. The *JV* curve and stabilized MPP tracking of the champion EDA-1 large area device (16.7% MPP PCE) are shown in **Figure S9**. We attribute this improvement to an improvement in the spin-coated film morphology over larger areas, which is crucial for any precursor solution to be suitable for scale-up and led us to investigate film morphology further.

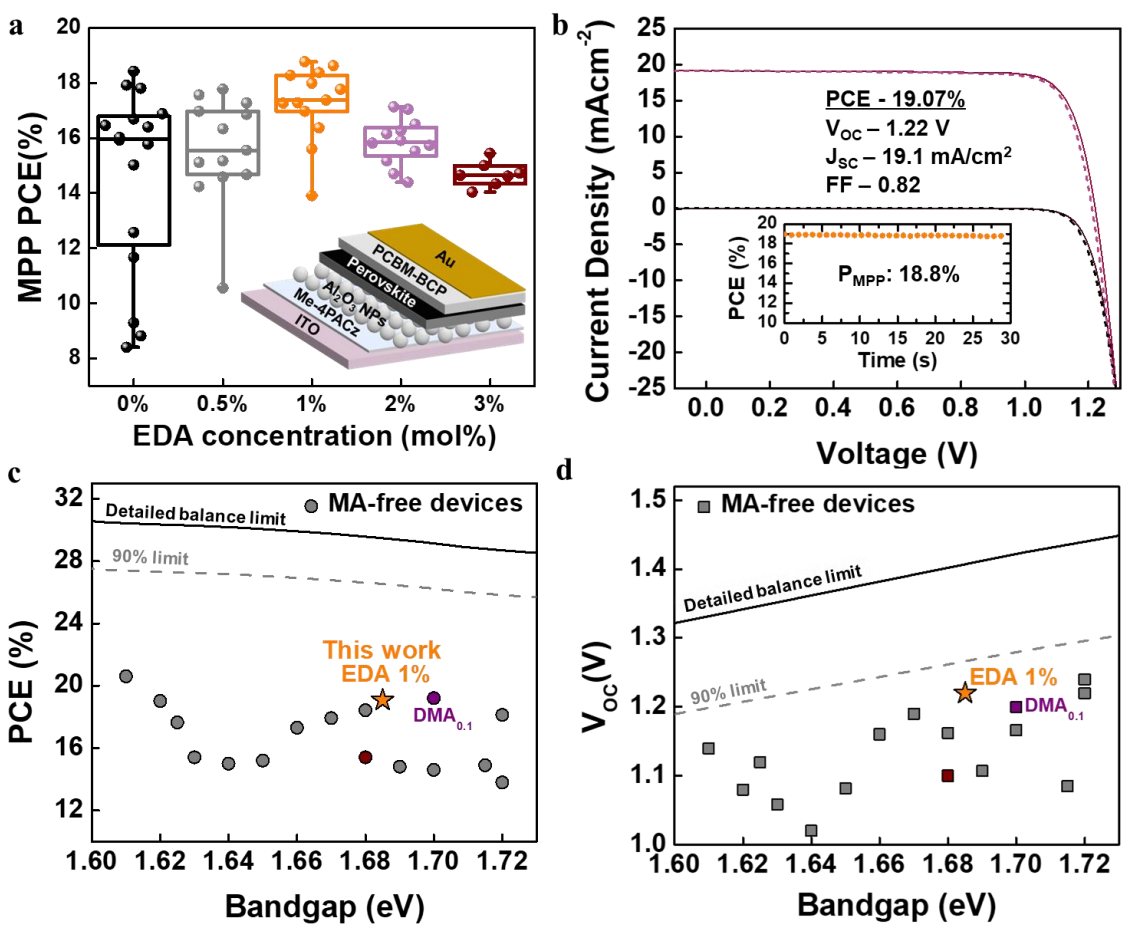

*Figure 2:* **Photovoltaic device results**. *(a) Inset: Schematic of the perovskite solar cell device architecture we employ. Main: efficiencies of solar cells made with 0-3 mol% EDA as an additive to the perovskite solution after 30 s of operation. (b) JV curves of the EDA-1 device under simulated solar illumination, recorded in forward (from short to open circuit) and reverse (from open to short circuit) bias. The inset tracks the PCE over 30 s of MPP-tracked device operation. (c) Power conversion efficiencies and (d) open circuit voltage ($V_{OC}$) results from the literature of wide gap devices made with $FA_{1-y}Cs_yPb(I_{1-x}Br_x)_3$ compositions. The additive used in the best performing devices are noted next to the $V_{OC}$ and PCE values. Purple data are from Ref.[23] using $DMA^+$. Brown points are from Ref.[36] using the same perovskite composition as this work ($FA_{0.83}Cs_{0.17}Pb(I_{0.75}Br_{0.25})_3$). All references used in parts (c) and (d) are given in **Table S2**.*

To better understand the origins of the improved stability, PL, and device performance, we explored how the PL varied on the microscale using hyperspectral microscopy (**Figure 3a and 3c**). A hyperspectral microscope obtains a PL spectrum at every pixel with high spatial resolution (for setup details see methods section) and has been used to spatially resolve heterogeneity in perovskites[37–39] and other semiconductors. **Figure 3b and 3d** plot summaries of the hyperspectral data cubes showing the wavelength-integrated total PL intensity, as well as the peak emission wavelength at every point in the sample. These measurements show that the that the EDA-1 processed films are significantly more homogeneous in both PL intensity, and PL peak wavelength than the EDA-0 films. The luminescent regions in EDA-0 are more red-shifted in accordance with them being more iodide-rich - narrower gap - regions into which charge carriers are funneled and recombine radiatively with high efficiency.[11,40] Generally, these microscopic results provide insight into the area-averaged spectra shown in Figure 1. Indeed, the histogram in **Figure 3b** shows the reference films exhibit a much wider distribution of local PL intensities, with many areas of the film being effectively dark. Similarly, the peak emission wavelength is also affected by the additive. Whereas the local emission peak wavelength varies between ~720 to 780 nm in the reference films, the EDA-1 films show a narrow clustering of the emission peaking around ~744 nm, consistent with the area-averaged data (**Figure 3d**). These data help explain the improved stability and device performance. The increased PLQY indicates a reduction in trap density in the EDA-1 films, facilitating larger quasi-Fermi-level splitting under illumination and therefore increased $V_{OC}$,[41–43] and additionally the improved homogeneity could help improve $V_{OC}$ of the EDA-1 films by reducing pinning to the low bandgap regions of the film, which is a common problem in complex semiconductors.[44,45] Moreover, it has been shown recently that compositional inhomogeneity of the perovskite phase is primary cause of phase instability in $FA_{1-x}Cs_{0.x}Pb(I_{1-}$

$_y$Br$_y$)$_3$ perovskites.[46–48] **Figure S10** shows the formation of "wrinkled" [49–52] films when using a higher precursor solution concentration (1.45 M). Even in this case, EDA-1 films show more homogeneous PL intensity and wavelength distributions compared to EDA-0. **Figure S11** also shows 10 mol% addition of EDA (EDA-10) shows homogenized PL and emission wavelength distribution in the microscale. To further investigate the compositional variation between the samples, we performed scanning electron microscopy energy-dispersive X-ray analysis (SEM EDS) analysis at various points over the film surfaces for the EDA-0 and EDA-10 films. **Figure S12** shows that the EDA-0 film has greater variation in atomic composition compared to the narrowly distributed EDA-10 film.

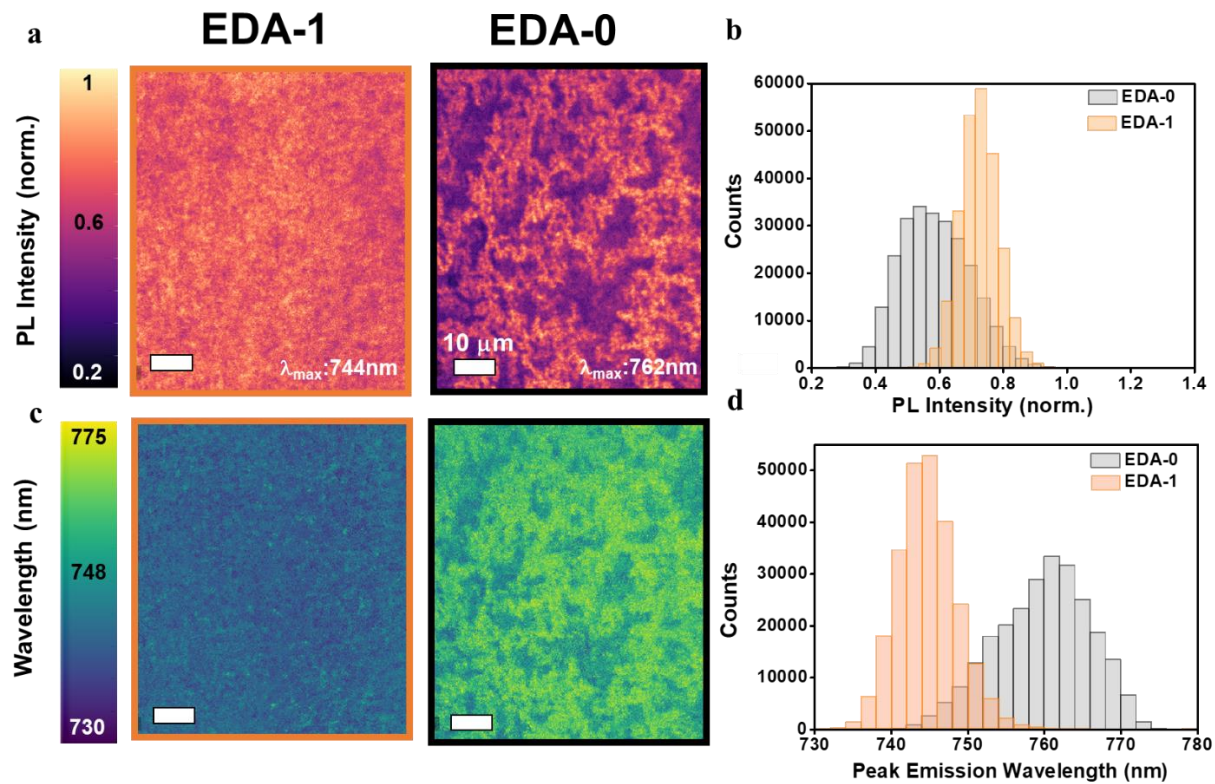

*Figure 3:* **Halide heterogeneity on the microscale.** *(a) Hyperspectral PL intensity map of the EDA-0 (black box) and EDA-1 (orange box). PL images are shown at the maximum emission*

*wavelength ($\lambda_{max}$) for each sample as stated on the figure (EDA-0 $\lambda_{max}$ = 762 nm, EDA-1 $\lambda_{max}$ = 744 nm). Excitation was achieved with a white lamp with short pass filter at 500 nm. The emission was recorded using a 500 nm dichroic and 650 nm long pass filter. (b) Histogram of the PL intensity distribution at the maximum emission wavelength for EDA-0 (grey) and EDA-1 (orange) samples. (c) Peak emission wavelength map of the EDA-0 and EDA-1 films. (d) Histogram of the distribution of peak emission wavelengths for the EDA-0 (grey) and EDA-1 (orange) samples.*

To better understand the effects of EDA addition on film formation, we investigated the crystallization of films by blade coating solutions and acquiring *in situ* grazing-incidence wide-angle X-ray scattering (GIWAXS) measurements (described in the Methods, scheme shown in **Figure S13**). For all samples (0%, 1% and 10% EDA) we identify that there exists an intermediate phase, which is a 2H polytype.[53] In the EDA-0 sample, this phase subsequently converts fully into a pseudo-cubic (3C) perovskite phase upon annealing (**Figure S14a**). However, with EDA-10 we find that in addition to the corner-sharing (α) perovskite phase, a different 2H polytype and a 4H polytype[53,54] are also formed in the fully annealed films (**Figure S14c**). For the EDA-1 sample, we detect a weak signature of a new polytype (**Figure S14f**). These results indicate that the EDA additive is altering the growth and resultant film composition at the studied concentrations.

In the SI we investigate and discuss the impact of higher EDA concentrations upon crystallization and phase behavior in thin films (**SI Note 1**). Significantly, we note that the XRD peaks from the 2H phase at a high EDA concentration (40 mol%) are a close match for a reported imidazolium lead iodide (ImPbI$_3$) 2H phase **(Figure S19)**.[55,56] We therefore investigated the possibility of larger cations forming in solution by conducting a range of NMR measurements. In **Figure 4a**, we show $^1$H NMR spectra of our perovskite precursor materials with and without 10 mol% EDA and compare these to neat FAI and EDA, with all materials dissolved in d$^6$-DMSO

(full spectra in **Figure S24**). Unexpectedly, the addition of 10 mol% EDA does not lead to the appearance of the signals observed for neat EDA (δ 2.47, 1.23 ppm) but instead to the appearance of two new $^1$H signals at 7.85 ppm (1H, t) and 3.62 ppm (4H, s). $^1$H-$^1$H correlation spectroscopy (COSY) reveals off-diagonal signals demonstrating spin-spin coupling between these two new signals, indicating that they correspond to $^1$H environments in the same molecule (**Figure 4b**). In $^{13}$C NMR (**Figure S25**) signals observed in spectra acquired from EDA, FAI and perovskite solutions correspond to the expected $^{13}$C environments. However, in the EDA-10 precursor solution we again observe two unexpected $^{13}$C signals, and the absence of a signal corresponding to the carbon environment of neat EDA. Significantly, $^1$H-$^{13}$C heteronuclear single quantum correlation (HSQC) spectroscopy of the EDA-10 precursor solution (**Figure S26**) reveals that the proton giving rise to the unexpected signal at 3.62 ppm is bonded to the carbon corresponding to the signal observed at 48.6 ppm, suggesting the presence of a new alkyl carbon bonded to an electronegative atom. The proximity of the new $^1$H and $^{13}$C signals at 7.85 ppm and 161.9 ppm, respectively, to those corresponding to the methine environment of $FA^+$ suggest that the new species formed in solution also contains an amidinium group. That the $^1$H signal at 3.62 ppm is shown to be both attached to an alkyl carbon and spin-spin coupled to the new $^1$H methine environment strongly implies the formation of a new species in solution by direct reaction between $FA^+$ and EDA. The 4:1 stoichiometry of alkyl:methine $^1$H signals suggests a 1:1 reaction. As the four alkyl hydrogens of EDA remain equivalent in the new species, we deduce that the symmetry of both $FA^+$ and EDA is retained in the new species.

Given these findings, we propose that consecutive addition-elimination reactions occur in solution leading to the formation of a secondary amidinium cation contained within a five-membered ring, 2-imidazolinium (4,5-dihydroimidazolium, $C_3H_7N_2^+$, $Imn^+$), and the elimination of two equivalents of volatile ammonia (inset of **Figure 4b**). We account mechanistically for this reaction in **Figure S27**. Such amine-amidinium reactions have precedent in recent perovskite literature,[57,58] and specifically this reaction is noted in early synthesis protocols for imidazolines.[59] To further support the proposed formation of Imn+, time-of-flight secondary ion mass spectrometry (ToF-SIMS) (**Figure S28**) data shows a high intensity $C_3H_7N_2^+$ mass fragment consistent with the formation of a large amount of $Imn^+$ in the film fabricated with 20 mol% EDA.

With this reaction we can rationalize why, in our study, the EDA additive does not result in a hollow perovskite, as in previous works EDA was generally employed in acidic growth conditions and so is easily protonated. Under such conditions, or in solutions where $EDA^{2+}$ is added directly, there is only a negligible quantity of ethylenediamine present at any time. As the diamine nucleophile is necessary to produce $Imn^+$ via reaction with $FA^+$, such conditions would effectively prevent its formation. To support this hypothesis, we verified that we do not form any polytype phases when we add the diammonium salts $EDAH_2I_2$ or $EDAH_2Br_2$ (**Figure S29**), or when we only have $MA^+$ rather than $FA^+$ in solution (**Figure S30**).

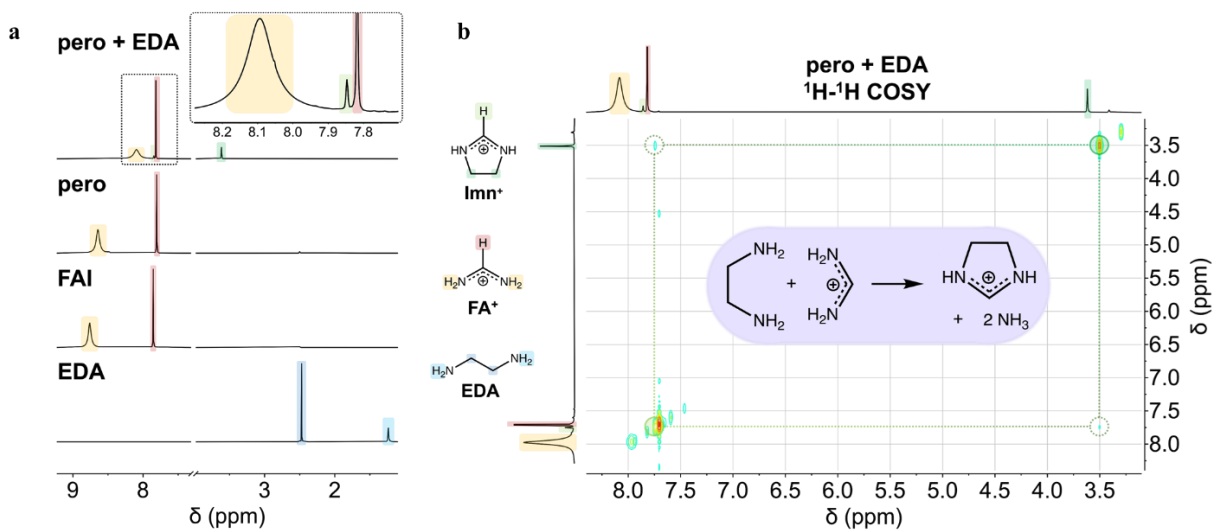

*Figure 4:* **Chemical Origins of EDA Effects** *(a) $^1$H solution NMR spectra of (bottom to top) ethylenediamine (EDA), formamidinium iodide and (FA$_{0.83}$Cs$_{0.17}$)Pb(I$_{0.75}$Br$_{0.25}$)$_3$ without and with 10 mol% EDA added. (b) $^1$H-$^1$H correlation spectroscopy (COSY) of (FA$_{0.83}$Cs$_{0.17}$)Pb(I$_{0.75}$Br$_{0.25}$)$_3$ with 10 mol% EDA added. Off-diagonal correlation indicates corresponding $^1$H chemical environments are spin-spin coupled. All materials are dissolved in d$^6$ DMSO. The inset highlighted in purple shows the overall reaction of FA$^+$ with EDA to produce Imn$^+$ and ammonia.*

In summary, we see that EDA has a beneficial effect on the stability of MA$^+$-free mixed-cation mixed-halide perovskite compositions. We are able to achieve an outstanding V$_{OC}$ and MPP-tracked PCE of 1.22 V and 18.8%, respectively, for MA$^+$-free devices with a bandgap of ~1.7 eV (1.69 eV), which is highly suitable for perovskite-on-Si tandem devices. We observe that EDA reacts with FA$^+$ in solution to form imidazolinium, which impacts the thin-film crystallization pathway and correlates with significantly improved compositional homogeneity, thin-film ambient stability and device performance. Importantly, this approach of using amine additives can be generalized to the production of other amidinium-derived compounds and may enable a range of species which are otherwise challenging to isolate. Similar approaches could be used to further engineer the surface chemistry and enhance the stability of halide perovskites.

**Acknowledgements**

This paper is based on work supported primarily by the Office of Naval Research (Award # N00014-20-1-2587). ToF-SIMS and part of the XRD analysis were carried out at the Molecular Analysis Facility, a National Nanotechnology Coordinated Infrastructure site at the University of Washington which is supported in part by the National Science Foundation (NNCI-1542101), the Molecular Engineering & Sciences Institute, and the Clean Energy Institute. The SEM analysis was carried out in the UW Clean Energy Institute Research Training Testbed facility. TRPL and some PL measurements were performed using the shared user facilities of UW's Molecular Engineering Materials Center, an NSF MRSEC (DMR-1719797). M.T. thanks Dan Graham at the Molecular Analysis Facility at the University of Washington for his help with ToF-SIMS data collection and analysis and Hannah Contreras for experimental help. D.S.G. acknowledges salary and infrastructure support from the Washington Research Foundation, the Alvin L. and Verla R. Kwiram endowment, and the B. Seymour Rabinovitch Endowment. S.Z. acknowledges receiving funding from the Rank Prize Funds. We acknowledge Diamond Light Source for time on beamline I07 under proposal SI30612-1 for the GIWAXS measurements and thank Saqlain Choudhary and Dr. Hadeel Hussain for assistance and Dr. Daniel Toolan and Karl-Augustin Zaininger for use of the *in situ* blade coater during these experiments.

**References**

(1) Correa-Baena, J.-P.; Saliba, M.; Buonassisi, T.; Grätzel, M.; Abate, A.; Tress, W.; Hagfeldt, A. Promises and Challenges of Perovskite Solar Cells. *Science* **2017**, *358* (6364), 739–744.

(2) Lin, K.; Xing, J.; Quan, L. N.; de Arquer, F. P. G.; Gong, X.; Lu, J.; Xie, L.; Zhao, W.; Zhang, D.; Yan, C.; Li, W.; Liu, X.; Lu, Y.; Kirman, J.; Sargent, E. H.; Xiong, Q.; Wei, Z. Perovskite Light-Emitting Diodes with External Quantum Efficiency Exceeding 20 per Cent. *Nature* **2018**, *562* (7726), 245–248.

(3) Basiricò, L.; Ciavatti, A.; Fraboni, B. Solution-Grown Organic and Perovskite X-Ray Detectors: A New Paradigm for the Direct Detection of Ionizing Radiation. *Advanced Materials Technologies* **2021**, *6* (1), 2000475.

(4) Muckel, F.; Guye, K. N.; Gallagher, S. M.; Liu, Y.; Ginger, D. S. Tuning Hybrid Exciton–Photon Fano Resonances in Two-Dimensional Organic–Inorganic Perovskite Thin Films. *Nano Lett.* **2021**, *21* (14), 6124–6131.

(5) Li, C.; Song, Z.; Chen, C.; Xiao, C.; Subedi, B.; Harvey, S. P.; Shrestha, N.; Subedi, K. K.; Chen, L.; Liu, D.; Li, Y.; Kim, Y.-W.; Jiang, C.; Heben, M. J.; Zhao, D.; Ellingson, R. J.; Podraza, N. J.; Al-Jassim, M.; Yan, Y. Low-Bandgap Mixed Tin–Lead Iodide Perovskites with Reduced Methylammonium for Simultaneous Enhancement of Solar Cell Efficiency and Stability. *Nat Energy* **2020**, *5* (10), 768–776.

(6)     Kulbak, M.; Cahen, D.; Hodes, G. How Important Is the Organic Part of Lead Halide Perovskite Photovoltaic Cells? Efficient CsPbBr$_3$ Cells. *J. Phys. Chem. Lett.* **2015**, *6* (13), 2452–2456.

(7)     McMeekin, D. P.; Sadoughi, G.; Rehman, W.; Eperon, G. E.; Saliba, M.; Horantner, M. T.; Haghighirad, A.; Sakai, N.; Korte, L.; Rech, B.; Johnston, M. B.; Herz, L. M.; Snaith, H. J. A Mixed-Cation Lead Mixed-Halide Perovskite Absorber for Tandem Solar Cells. *Science* **2016**, *351* (6269), 151–155.

(8)     Leijtens, T.; Bush, K. A.; Prasanna, R.; McGehee, M. D. Opportunities and Challenges for Tandem Solar Cells Using Metal Halide Perovskite Semiconductors. *Nat Energy* **2018**, *3* (10), 828–838.

(9)     Eperon, G. E.; Hörantner, M. T.; Snaith, H. J. Metal Halide Perovskite Tandem and Multiple-Junction Photovoltaics. *Nat Rev Chem* **2017**, *1* (12), 0095.

(10)    Lin, Y.-H.; Sakai, N.; Da, P.; Wu, J.; Sansom, H. C.; Ramadan, A. J.; Mahesh, S.; Liu, J.; Oliver, R. D. J.; Lim, J.; Aspitarte, L.; Sharma, K.; Madhu, P. K.; Morales-Vilches, A. B.; Nayak, P. K.; Bai, S.; Gao, F.; Grovenor, C. R. M.; Johnston, M. B.; Labram, J. G.; Durrant, J. R.; Ball, J. M.; Wenger, B.; Stannowski, B.; Snaith, H. J. A Piperidinium Salt Stabilizes Efficient Metal-Halide Perovskite Solar Cells. *Science* **2020**, *369* (6499), 96–102.

(11)    Motti, S. G.; Patel, J. B.; Oliver, R. D. J.; Snaith, H. J.; Johnston, M. B.; Herz, L. M. Phase Segregation in Mixed-Halide Perovskites Affects Charge-Carrier Dynamics While Preserving Mobility. *Nat Commun* **2021**, *12* (1), 6955.

(12)    Wieghold, S.; Tresback, J.; Correa-Baena, J.-P.; Hartono, N. T. P.; Sun, S.; Liu, Z.; Layurova, M.; VanOrman, Z. A.; Bieber, A. S.; Thapa, J.; Lai, B.; Cai, Z.; Nienhaus, L.; Buonassisi, T. Halide Heterogeneity Affects Local Charge Carrier Dynamics in Mixed-Ion Lead Perovskite Thin Films. *Chem. Mater.* **2019**, *31* (10), 3712–3721.

(13)    Juarez-Perez, E. J.; Ono, L. K.; Maeda, M.; Jiang, Y.; Hawash, Z.; Qi, Y. Photodecomposition and Thermal Decomposition in Methylammonium Halide Lead Perovskites and Inferred Design Principles to Increase Photovoltaic Device Stability. *J. Mater. Chem. A* **2018**, *6* (20), 9604–9612.

(14)    Conings, B.; Drijkoningen, J.; Gauquelin, N.; Babayigit, A.; D'Haen, J.; D'Olieslaeger, L.; Ethirajan, A.; Verbeeck, J.; Manca, J.; Mosconi, E.; Angelis, F. D.; Boyen, H.-G. Intrinsic Thermal Instability of Methylammonium Lead Trihalide Perovskite. *Advanced Energy Materials* **2015**, *5* (15), 1500477.

(15)    An, Y.; Hidalgo, J.; Perini, C. A. R.; Castro-Méndez, A.-F.; Vagott, J. N.; Bairley, K.; Wang, S.; Li, X.; Correa-Baena, J.-P. Structural Stability of Formamidinium- and Cesium-Based Halide Perovskites. *ACS Energy Lett.* **2021**, *6* (5), 1942–1969.

(16)    Kerner, R. A.; Schloemer, T. H.; Schulz, P.; Berry, J. J.; Schwartz, J.; Sellinger, A.; Rand, B. P. Amine Additive Reactions Induced by the Soft Lewis Acidity of Pb$^{2+}$ in Halide

Perovskites. Part II: Impacts of Amido Pb Impurities in Methylammonium Lead Triiodide Thin Films. *J. Mater. Chem. C* **2019**, *7* (18), 5244–5250.

(17)     Kerner, R. A.; Schloemer, T. H.; Schulz, P.; Berry, J. J.; Schwartz, J.; Sellinger, A.; Rand, B. P. Amine Additive Reactions Induced by the Soft Lewis Acidity of $Pb^{2+}$ in Halide Perovskites. Part I: Evidence for Pb–Alkylamide Formation. *J. Mater. Chem. C* **2019**, *7* (18), 5251–5259.

(18)     Wu, W.-Q.; Yang, Z.; Rudd, P. N.; Shao, Y.; Dai, X.; Wei, H.; Zhao, J.; Fang, Y.; Wang, Q.; Liu, Y.; Deng, Y.; Xiao, X.; Feng, Y.; Huang, J. Bilateral Alkylamine for Suppressing Charge Recombination and Improving Stability in Blade-Coated Perovskite Solar Cells. *Science Advances 5* (3), eaav8925.

(19)     Xu, Y.; Xu, W.; Hu, Z.; Steele, J. A.; Wang, Y.; Zhang, R.; Zheng, G.; Li, X.; Wang, H.; Zhang, X.; Solano, E.; Roeffaers, M. B. J.; Uvdal, K.; Qing, J.; Zhang, W.; Gao, F. Impact of Amine Additives on Perovskite Precursor Aging: A Case Study of Light-Emitting Diodes. *J. Phys. Chem. Lett.* **2021**, *12* (25), 5836–5843.

(20)     Ke, W.; Stoumpos, C. C.; Zhu, M.; Mao, L.; Spanopoulos, I.; Liu, J.; Kontsevoi, O. Y.; Chen, M.; Sarma, D.; Zhang, Y.; Wasielewski, M. R.; Kanatzidis, M. G. Enhanced Photovoltaic Performance and Stability with a New Type of Hollow 3D Perovskite {en}$FASnI_3$. *Sci. Adv.* **2017**, *3* (8), e1701293.

(21)     Li, C.; Guerrero, A.; Huettner, S.; Bisquert, J. Unravelling the Role of Vacancies in Lead Halide Perovskite through Electrical Switching of Photoluminescence. *Nature Communications* **2018**, *9* (1), 5113.

(22)     Hu, S.; Otsuka, K.; Murdey, R.; Nakamura, T.; Truong, M. A.; Yamada, T.; Handa, T.; Matsuda, K.; Nakano, K.; Sato, A.; Marumoto, K.; Tajima, K.; Kanemitsu, Y.; Wakamiya, A. Optimized Carrier Extraction at Interfaces for 23.6% Efficient Tin–Lead Perovskite Solar Cells. *Energy Environ. Sci.* **2022**, *15* (5), 2096–2107.

(23)     Palmstrom, A. F.; Eperon, G. E.; Leijtens, T.; Prasanna, R.; Habisreutinger, S. N.; Nemeth, W.; Gaulding, E. A.; Dunfield, S. P.; Reese, M.; Nanayakkara, S.; Moot, T.; Werner, J.; Liu, J.; To, B.; Christensen, S. T.; McGehee, M. D.; van Hest, M. F. A. M.; Luther, J. M.; Berry, J. J.; Moore, D. T. Enabling Flexible All-Perovskite Tandem Solar Cells. *Joule* **2019**, *3* (9), 2193–2204.

(24)     Eperon, G. E.; Stone, K. H.; Mundt, L. E.; Schloemer, T. H.; Habisreutinger, S. N.; Dunfield, S. P.; Schelhas, L. T.; Berry, J. J.; Moore, D. T. The Role of Dimethylammonium in Bandgap Modulation for Stable Halide Perovskites. *ACS Energy Lett.* **2020**, *5* (6), 1856–1864.

(25)     Jariwala, S.; Kumar, R. E.; Eperon, G. E.; Shi, Y.; Fenning, D. P.; Ginger, D. S. Dimethylammonium Addition to Halide Perovskite Precursor Increases Vertical and Lateral Heterogeneity. *ACS Energy Lett.* **2022**, *7* (1), 204–210.

(26) Yoon, S. J.; Kuno, M.; Kamat, P. V. *Shift Happens* . How Halide Ion Defects Influence Photoinduced Segregation in Mixed Halide Perovskites. *ACS Energy Lett.* **2017**, *2* (7), 1507–1514.

(27) Slotcavage, D. J.; Karunadasa, H. I.; McGehee, M. D. Light-Induced Phase Segregation in Halide-Perovskite Absorbers. *ACS Energy Lett.* **2016**, *1* (6), 1199–1205.

(28) Barker, A. J.; Sadhanala, A.; Deschler, F.; Gandini, M.; Senanayak, S. P.; Pearce, P. M.; Mosconi, E.; Pearson, A. J.; Wu, Y.; Srimath Kandada, A. R.; Leijtens, T.; De Angelis, F.; Dutton, S. E.; Petrozza, A.; Friend, R. H. Defect-Assisted Photoinduced Halide Segregation in Mixed-Halide Perovskite Thin Films. *ACS Energy Lett.* **2017**, *2* (6), 1416–1424.

(29) Oliver, R. D. J.; Caprioglio, P.; Peña-Camargo, F.; Buizza, L. R. V.; Zu, F.; Ramadan, A. J.; Motti, S. G.; Mahesh, S.; McCarthy, M. M.; Warby, J. H.; Lin, Y.-H.; Koch, N.; Albrecht, S.; Herz, L. M.; Johnston, M. B.; Neher, D.; Stolterfoht, M.; Snaith, H. J. Understanding and Suppressing Non-Radiative Losses in Methylammonium-Free Wide-Bandgap Perovskite Solar Cells. *Energy Environ. Sci.* **2022**, 10.1039.D1EE02650J.

(30) Jariwala, S.; Burke, S.; Dunfield, S.; Shallcross, R. C.; Taddei, M.; Wang, J.; Eperon, G. E.; Armstrong, N. R.; Berry, J. J.; Ginger, D. S. Reducing Surface Recombination Velocity of Methylammonium-Free Mixed-Cation Mixed-Halide Perovskites via Surface Passivation. *Chem. Mater.* **2021**, acs.chemmater.1c00848.

(31) Khenkin, M. V.; Katz, E. A.; Abate, A.; Bardizza, G.; Berry, J. J.; Brabec, C.; Brunetti, F.; Bulović, V.; Burlingame, Q.; Di Carlo, A.; Cheacharoen, R.; Cheng, Y.-B.; Colsmann, A.; Cros, S.; Domanski, K.; Dusza, M.; Fell, C. J.; Forrest, S. R.; Galagan, Y.; Di Girolamo, D.; Grätzel, M.; Hagfeldt, A.; von Hauff, E.; Hoppe, H.; Kettle, J.; Köbler, H.; Leite, M. S.; Liu, S. (Frank); Loo, Y.-L.; Luther, J. M.; Ma, C.-Q.; Madsen, M.; Manceau, M.; Matheron, M.; McGehee, M.; Meitzner, R.; Nazeeruddin, M. K.; Nogueira, A. F.; Odabaşı, Ç.; Osherov, A.; Park, N.-G.; Reese, M. O.; De Rossi, F.; Saliba, M.; Schubert, U. S.; Snaith, H. J.; Stranks, S. D.; Tress, W.; Troshin, P. A.; Turkovic, V.; Veenstra, S.; Visoly-Fisher, I.; Walsh, A.; Watson, T.; Xie, H.; Yıldırım, R.; Zakeeruddin, S. M.; Zhu, K.; Lira-Cantu, M. Consensus Statement for Stability Assessment and Reporting for Perovskite Photovoltaics Based on ISOS Procedures. *Nat Energy* **2020**, *5* (1), 35–49.

(32) Eperon, G. E.; Paternò, G. M.; Sutton, R. J.; Zampetti, A.; Haghighirad, A. A.; Cacialli, F.; Snaith, H. J. Inorganic Caesium Lead Iodide Perovskite Solar Cells. *J. Mater. Chem. A* **2015**, *3* (39), 19688–19695.

(33) Hu, Y.; Aygüler, M. F.; Petrus, M. L.; Bein, T.; Docampo, P. Impact of Rubidium and Cesium Cations on the Moisture Stability of Multiple-Cation Mixed-Halide Perovskites. *ACS Energy Lett.* **2017**, *2* (10), 2212–2218.

(34) Al-Ashouri, A.; Köhnen, E.; Li, B.; Magomedov, A.; Hempel, H.; Caprioglio, P.; Márquez, J. A.; Morales Vilches, A. B.; Kasparavicius, E.; Smith, J. A.; Phung, N.; Menzel, D.; Grischek, M.; Kegelmann, L.; Skroblin, D.; Gollwitzer, C.; Malinauskas, T.; Jošt, M.; Matič, G.; Rech, B.; Schlatmann, R.; Topič, M.; Korte, L.; Abate, A.; Stannowski, B.; Neher, D.;

Stolterfoht, M.; Unold, T.; Getautis, V.; Albrecht, S. Monolithic Perovskite/Silicon Tandem Solar Cell with >29% Efficiency by Enhanced Hole Extraction. *Science* **2020**, *370* (6522), 1300–1309.

(35) Rakocevic, L.; Ernst, F.; Yimga, N. T.; Vashishtha, S.; Aernouts, T.; Heumueller, T.; Brabec, C. J.; Gehlhaar, R.; Poortmans, J. Reliable Performance Comparison of Perovskite Solar Cells Using Optimized Maximum Power Point Tracking. *Solar RRL* **2019**, *3* (2), 1800287.

(36) Bush, K. A.; Frohna, K.; Prasanna, R.; Beal, R. E.; Leijtens, T.; Swifter, S. A.; McGehee, M. D. Compositional Engineering for Efficient Wide Band Gap Perovskites with Improved Stability to Photoinduced Phase Segregation. *ACS Energy Lett.* **2018**, *3* (2), 428–435.

(37) Frohna, K.; Anaya, M.; Macpherson, S.; Sung, J.; Doherty, T. A. S.; Chiang, Y.-H.; Winchester, A. J.; Orr, K. W. P.; Parker, J. E.; Quinn, P. D.; Dani, K. M.; Rao, A.; Stranks, S. D. Nanoscale Chemical Heterogeneity Dominates the Optoelectronic Response of Alloyed Perovskite Solar Cells. *Nat. Nanotechnol.* **2022**, *17* (2), 190–196.

(38) Tennyson, E. M.; Frohna, K.; Drake, W. K.; Sahli, F.; Chien-Jen Yang, T.; Fu, F.; Werner, J.; Chosy, C.; Bowman, A. R.; Doherty, T. A. S.; Jeangros, Q.; Ballif, C.; Stranks, S. D. Multimodal Microscale Imaging of Textured Perovskite–Silicon Tandem Solar Cells. *ACS Energy Lett.* **2021**, 2293–2304.

(39) El-Hajje, G.; Momblona, C.; Gil-Escrig, L.; Ávila, J.; Guillemot, T.; Guillemoles, J.-F.; Sessolo, M.; Bolink, H. J.; Lombez, L. Quantification of Spatial Inhomogeneity in Perovskite Solar Cells by Hyperspectral Luminescence Imaging. *Energy Environ. Sci.* **2016**, *9* (7), 2286–2294.

(40) Feldmann, S.; Macpherson, S.; Senanayak, S. P.; Abdi-Jalebi, M.; Rivett, J. P. H.; Nan, G.; Tainter, G. D.; Doherty, T. A. S.; Frohna, K.; Ringe, E.; Friend, R. H.; Sirringhaus, H.; Saliba, M.; Beljonne, D.; Stranks, S. D.; Deschler, F. Photodoping through Local Charge Carrier Accumulation in Alloyed Hybrid Perovskites for Highly Efficient Luminescence. *Nat. Photonics* **2020**, *14* (2), 123–128.

(41) Yao, J.; Kirchartz, T.; Vezie, M. S.; Faist, M. A.; Gong, W.; He, Z.; Wu, H.; Troughton, J.; Watson, T.; Bryant, D.; Nelson, J. Quantifying Losses in Open-Circuit Voltage in Solution-Processable Solar Cells. *Phys. Rev. Applied* **2015**, *4* (1), 014020.

(42) Miller, O. D.; Yablonovitch, E.; Kurtz, S. R. Strong Internal and External Luminescence as Solar Cells Approach the Shockley-Queisser Limit. *IEEE J. Photovolt* **2012**, 303–311.

(43) Braly, I. L.; deQuilettes, D. W.; Pazos-Outón, L. M.; Burke, S.; Ziffer, M. E.; Ginger, D. S.; Hillhouse, H. W. Hybrid Perovskite Films Approaching the Radiative Limit with over 90% Photoluminescence Quantum Efficiency. *Nature Photon* **2018**, *12* (6), 355–361.

(44) Stanbery, B. J. Copper Indium Selenides and Related Materials for Photovoltaic Devices. *Critical Reviews in Solid State and Materials Sciences* **2002**, *27* (2), 73–117.

(45) Wilson, G. M.; Al-Jassim, M.; Metzger, W. K.; Glunz, S. W.; Verlinden, P.; Xiong, G.; Mansfield, L. M.; Stanbery, B. J.; Zhu, K.; Yan, Y.; Berry, J. J.; Ptak, A. J.; Dimroth, F.; Kayes, B. M.; Tamboli, A. C.; Peibst, R.; Catchpole, K.; Reese, M. O.; Klinga, C. S.; Denholm, P.; Morjaria, M.; Deceglie, M. G.; Freeman, J. M.; Mikofski, M. A.; Jordan, D. C.; TamizhMani, G.; Sulas-Kern, D. B. The 2020 Photovoltaic Technologies Roadmap. *J. Phys. D: Appl. Phys.* **2020**, *53* (49), 493001

(46) Barrier, J.; Beal, R. E.; Gold-Parker, A.; Vigil, J. A.; Wolf, E.; Waquier, L.; Weadock, N. J.; Zhang, Z.; Schelhas, L. T.; Nogueira, A. F.; McGehee, M. D.; Toney, M. F. Compositional Heterogeneity in $Cs_y FA_{1-y} Pb(Br_x I_{1-x})_3$ Perovskite Films and Its Impact on Phase Behavior. *Energy Environ. Sci.* **2021**, *14* (12), 6394–6405.

(47) Mundt, L. E.; Zhang, F.; Palmstrom, A. F.; Xu, J.; Tirawat, R.; Kelly, L. L.; Stone, K. H.; Zhu, K.; Berry, J. J.; Toney, M. F.; Schelhas, L. T. Mixing Matters: Nanoscale Heterogeneity and Stability in Metal Halide Perovskite Solar Cells. *ACS Energy Lett.* **2022**, *7* (1), 471–480.

(48) Macpherson, S.; Doherty, T. A. S.; Winchester, A. J.; Kosar, S.; Johnstone, D. N.; Chiang, Y.-H.; Galkowski, K.; Anaya, M.; Frohna, K.; Iqbal, A. N.; Nagane, S.; Roose, B.; Andaji-Garmaroudi, Z.; Orr, K. W. P.; Parker, J. E.; Midgley, P. A.; Dani, K. M.; Stranks, S. D. Local Nanoscale Phase Impurities Are Degradation Sites in Halide Perovskites. *Nature* **2022**, *607* (7918), 294–300.

(49) Bercegol, A.; Ramos, F. J.; Rebai, A.; Guillemot, T.; Puel, J.-B.; Guillemoles, J.-F.; Ory, D.; Rousset, J.; Lombez, L. Spatial Inhomogeneity Analysis of Cesium-Rich Wrinkles in Triple-Cation Perovskite. *J. Phys. Chem. C* **2018**, *122* (41), 23345–23351.

(50) Braunger, S.; Mundt, L. E.; Wolff, C. M.; Mews, M.; Rehermann, C.; Jošt, M.; Tejada, A.; Eisenhauer, D.; Becker, C.; Guerra, J. A.; Unger, E.; Korte, L.; Neher, D.; Schubert, M. C.; Rech, B.; Albrecht, S. $Cs_x FA_{1-x} Pb(I_{1-y} Br_y)_3$ Perovskite Compositions: The Appearance of Wrinkled Morphology and Its Impact on Solar Cell Performance. *J. Phys. Chem. C* **2018**, *122* (30), 17123–17135.

(51) Bush, K. A.; Rolston, N.; Gold-Parker, A.; Manzoor, S.; Hausele, J.; Yu, Z. J.; Raiford, J. A.; Cheacharoen, R.; Holman, Z. C.; Toney, M. F.; Dauskardt, R. H.; McGehee, M. D. Controlling Thin-Film Stress and Wrinkling during Perovskite Film Formation. *ACS Energy Lett.* **2018**, *3* (6), 1225–1232.

(52) Kim, S.-G.; Kim, J.-H.; Ramming, P.; Zhong, Y.; Schötz, K.; Kwon, S. J.; Huettner, S.; Panzer, F.; Park, N.-G. How Antisolvent Miscibility Affects Perovskite Film Wrinkling and Photovoltaic Properties. *Nat Commun* **2021**, *12* (1), 1554.

(53) Gratia, P.; Zimmermann, I.; Schouwink, P.; Yum, J.-H.; Audinot, J.-N.; Sivula, K.; Wirtz, T.; Nazeeruddin, M. K. The Many Faces of Mixed Ion Perovskites: Unraveling and Understanding the Crystallization Process. *ACS Energy Lett.* **2017**, *2* (12), 2686–2693.

(54)    Stoumpos, C. C.; Mao, L.; Malliakas, C. D.; Kanatzidis, M. G. Structure–Band Gap Relationships in Hexagonal Polytypes and Low-Dimensional Structures of Hybrid Tin Iodide Perovskites. *Inorg. Chem.* **2017**, *56* (1), 56–73.

(55)    Weber, O. J.; Marshall, K. L.; Dyson, L. M.; Weller, M. T. Structural Diversity in Hybrid Organic–Inorganic Lead Iodide Materials. *Acta Crystallographica Section B* **2015**, *71* (6), 668–678.

(56)    Seth, C.; Khushalani, D. Non-Perovskite Hybrid Material, Imidazolium Lead Iodide, with Enhanced Stability. *ChemNanoMat* **2019**, *5* (1), 85–91.

(57)    Li, Z. Ammonia for Post-Healing of Formamidinium-Based Perovskite Films. *Nature Communications* **2022**, 10.

(58)    Wang, X.; Fan, Y.; Wang, L.; Chen, C.; Li, Z.; Liu, R.; Meng, H.; Shao, Z.; Du, X.; Zhang, H.; Cui, G.; Pang, S. Perovskite Solution Aging: What Happened and How to Inhibit? *Chem* **2020**, *6* (6), 1369–1378.

(59)    Ferm, R. J.; Riebsomer, J. L. The Chemistry of the 2-Imidazolines and Imidazolidines. *Chem. Rev.* **1954**, *54* (4), 593–613.

# Supplementary Information

# Ethylenediamine Addition Improves Performance and Suppresses Phase Instabilities in Mixed-Halide Perovskites

Margherita Taddei,[1] Joel A. Smith,[2] Benjamin M. Gallant,[2] Suer Zhou,[2] Robert J. E. Westbrook,[1] Yangwei Shi,[1] Jian Wang,[1] James N. Drysdale,[2] Declan P. McCarthy,[3] Stephen Barlow,[3] Seth R. Marder,[3,4] Henry J. Snaith[2] and David S. Ginger[1]*

AUTHOR ADDRESS: Department of Chemistry, University of Washington, Seattle, WA, USA

AUTHOR INFORMATION

1 Department of Chemistry, University of Washington, Seattle WA, USA

2 Department of Physics, University of Oxford, Oxford, UK

3 Renewable and Sustainable Energy Institute, University of Colorado Boulder, Boulder CO, USA

4 Department of Chemical and Biological Engineering and Department of Chemistry, University of Colorado Boulder, Boulder CO, USA

**Corresponding Author**

* David S. Ginger, Department of Chemistry, University of Washington, Seattle WA, USA

**Experimental Methods**

*Materials:*

Films prepared at the University of Washington: Lead bromide (PbBr$_2$, 99.998% metals basis), and cesium iodide (CsI, 99.999% metal basis) were purchased from Alfa Aesar. Lead iodide (PbI$_2$, perovskite grade) was purchased from Tokyo Chemical Industries ltd. Formamidinium iodide (FAI, >99.99%) and methylammonium iodide (MAI, >99.99%) were purchased from Great Cell Solar. Ethylenediamine (puriss. p. a., absolute ≥99.5%), ethane-1,2-diammonium iodide (EDAH$_2$I$_2$), ethane-1,2-diammonium bromide (EDAH$_2$Br$_2$), lead acetate trihydrate (99.999% trace metals basis), dimethylformamide (DMF, anhydrous), dimethyl sulfoxide (DMSO, anhydrous), anisole (anhydrous) were purchased from Sigma-Aldrich. Films and devices prepared at the University of Oxford: Lead (II) bromide (PbBr$_2$ for perovskite precursor), lead (II) iodide (PbI$_2$ 99.99% trace metals basis for perovskite precursor) and Me-4PACz ([4-(3,6-dimethyl-9*H*-carbazol-9-yl)butyl]phosphonic acid) were purchased from Tokyo Chemical Industries ltd. BCP (bathocuproine, 98%) and cesium iodide (CsI, 99.9% trace metals basis) were purchased from Alfa Aesar. PCBM ([6,6]-phenyl-C$_{61}$-butyric acid methyl ester) was purchased from Solenne. Formamidinium iodide (FAI, 99.99%) was purchased from Dyenamo. Ethanol (anhydrous) was purchased from VWR. Remaining solvents and materials were purchased from Sigma-Aldrich.

*Synthesis:*

The synthesis of FA$_{0.83}$Cs$_{0.17}$Pb(I$_{1-x}$Br$_x$)$_3$ thin films was all conducted inside a N$_2$ filled glovebox. The FA$_{0.83}$Cs$_{0.17}$Pb(I$_{0.75}$Br$_{0.25}$)$_3$ precursor solutions of 1.2M were prepared dissolving 0.75 mmol of PbI$_2$, 0.45 mmol of PbBr$_2$, 0.204 mmol of CsI and 0.996 mmol of FAI in 0.8 mL DMF and 0.2 mL of DMSO. Solutions of 1.45M were prepared mixing 0.9 mmol of PbI$_2$, 0.54 mmol of PbBr$_2$, 0.25 mmol of CsI and 1.218 mmol of FAI in 0.8 mL DMF and 0.2 mL DMSO. All the solutions were stirred overnight at 600 rpm. After stirring, the solutions were filtered with a 0.2 µm PTFE filter. Ethylenediamine (EDA) was added to the precursor solution after overnight stirring in molar percentages of 1, 10, 20, 40 (density=0.899 g/mL). Solutions were deposited after a few hours after EDA addition. Glass substates (2.5 × 2.5 cm) were cleaned through a sequential sonication of 10 minutes in soap solution, DI water, acetone, and IPA. Cleaned glass substrates were ozone cleaned for 30 min before spin coating. The precursor solution (80 µL) was deposited on the glass substrates and spin coated for 10 s at 1000 rpm (200 rpm/s) and 35 s at 5000 rpm (800 rpm/s). Anisole (Sigma Aldrich) was used as antisolvent and dropped (330 uL) after 40 s. Anisole was filtered using a 0.45 µm PTFE filter before usage. The films were annealed at 100 °C for 45 min. Then the films were stored under $N_2$ in the dark. Films were encapsulated with a glass substrate (2.5 x 2.5 cm) using a 2-part epoxy resin-polymercaptan glue (Epoxy Adhesive C-POXY 30 by CECCORP) mixed in ratio 1:1. Encapsulation was performed inside the glove box. Encapsulated films were left inside the glove box overnight before characterization.

For films containing diammonium salts (**Figure S29**), $EDAH_2I_2$ and $EDAH_2Br_2$ were added to the precursor solution at 10 mol% excess, after which films were synthesized following the procedure listed above. For $MAPbI_3$ films shown in **Figure S30**, solutions and films were prepared following the procedure from Ref.[1]

*UV-Vis:*

UV-Vis absorbance spectra of the thin films were carried out on an Agilent 8453 UV-Vis Spectrometer in a range of 200-1100 nm and an integration time of 0.5 s. Tauc plots were extracted calculating the absorption coefficient α using equation:

$$\alpha = \frac{2.303 * A}{thickness} \tag{3}$$

Where A is the absorbance corrected for background scattering. All film thicknesses were estimated to be 500 nm for Tauc plotting.

*Photoluminescence:*

PL spectra were acquired using Edinburgh FLS1000 spectrometer. The excitation source is a 450 W ozone-free Xenon arc lamp and the detector a Si-PMT with spectral resolution from 200 to 980 nm. The films were characterized by using the instrument solid-state samples holder and the position was held constant through all the measured samples. The emission bandwidth was kept at 2 nm and dwell time at 0.3 s.

*Photoluminescence quantum yield (PLQY):*

PLQY results are determined using a 532 nm continuous wave laser (CrystaLaser LC) as the excitation source to illuminate the samples in an integrating sphere (Hamamatsu photonics K.K). The beam intensity was fixed equal to 1 Sun intensity. The power density was verified using an optical beam profiler. The PLQY is calculated following the formula:

$$PLQY = \frac{I_{em,sample} - I_{em.\ blank}}{I_{exc,blank} - I_{exc,sample}} * 100 \tag{4}$$

Where $I_{em,sample}$ and $I_{em,blank}$ are the integrated area under the curve in the emission region (650-950 nm) of the sample and the bare glass blank respectively. The $I_{exc,sample}$ and $I_{exc,blank}$ are the integrated area under the curve in the excitation region (450-630 nm) of the sample and the bare glass blank respectively.

*Time Resolved Photoluminescence:*

The TRPL spectra at 405 nm excitation were acquired using a Edinburgh FLS1000 spectrometer with EPL-405, a 405 nm picosecond pulsed diode laser. The repetition rate is controlled by an internal trigger input and was set to 1 MHz. The emission and excitation slits were controlled to have an emission signal frequency of 1 to 2% of the start rate. The slits were kept constants for all the measurements and the film stage was moved in the x direction to maximize the emission. A PMT-900 detector was used in TCSPC mode with an instrumental response width of approximately 600 ps.

The TRPL spectra at 640 nm excitation were acquired using a PicoQuant Picoharp 300 TCSPC system equipped with a 640 nm pulsed diode laser (PDL-800 LDH-P-C-470B, 300 ps pulse width). The laser was pulsed at repetition rates of 1 MHz. The PL emission was filtered using a 700 nm long-pass filter before being directed to the detector.

PL decay traces were fitted using a stretched-exponential decay function shown in **Eq.1** in the main text. This decay law is typically encountered in systems with a distribution of local decay rates and the β factor can give information pertaining to a random distribution of site energies or a time dependent rate constant. When β = 1 the decay function reduces to a single exponential and heterogeneity is negligible. When β is closer to 0, this represents a larger distribution (e.g. decay rates) and therefore more significant heterogeneity.[2] Using the experimentally measured characteristic lifetime and β values, we calculate the total distribution and obtain the average lifetime, $\tau_{stretch}$, using Eq.2. $\Gamma(\frac{1}{\beta})$ is defined as the gamma function:

$$\boldsymbol{\Gamma} = \int_0^\infty x^{\frac{(1-\beta)}{\beta}} e^{-x} dx \tag{5}$$

*Hyperspectral microscope:*

Hyperspectral measurements were performed using a Photon etc. IMA upright microscope fitted with a transmitted darkfield condenser and a 60X objective (Nikon Plan RT, NA 0.7, CC 0-1.2). The excitation was done using a mercury halide lamp passing through a 500 nm short-pass filter and emission was collected through a 500 dichroic filter and 550 nm long-pass filter. The Hyperspectral Microscope uses a tunable Bragg filter to image a sample at specific wavelengths.

The sample is imaged throughout the spectral range and these images are combined into a single "Hyper Cube", which carries spectral information at each pixel. This allows for diffraction limited imaging, rather than being constrained by the spot size of the fiber optic cable. Post-processing was done in the proprietary Photon etc. PHySpec Software.

*X-ray Diffraction:*

XRD patterns were obtained using a Bruker D8 Discover with IµS 2-D XRD System (Cu Kα radiation at 50 W). Further XRD at the University of Oxford were collected using a Panalytical X'Pert Pro operating at 40 mA and 40 kV accelerating voltage. Powder XRD patterns and structural views were calculated and rendered using VESTA.[3]

*Scanning Electron Microscopy:*

SEM images were obtained using a ThermoFisher Phenom ProX Desktop SEM with an integrated energy-dispersive X-ray diffraction (EDS) detector. Acceleration voltage of the electron beam was set to 15kV for EDS elemental analysis.

*ToF-SIMS:*

ToF-SIMS spectra were acquired on a IONTOF TOF.SIMS5 spectrometer using a 25 keV $Bi^{3+}$ cluster ion source in the pulsed mode. Spectra were acquired for both positive and negative secondary ions over a mass range of m/z = 0 to 800. The ion source was operated with at a current of 0.09 pA and spectra were acquired using an analysis area of 200 µm x 200 µm using 256 x 256 pixels.

*Device preparation:*

Me-4PACz solution was prepared based on the procedure of Ref.[4]. Briefly, a 1 mgml$^{-1}$ stock solution in anhydrous ethanol was sonicated for ~15 mins and stirred overnight, then filtered with a 0.45 µm PTFE filter. This stock solution was then diluted to 0.33 mgml$^{-1}$ (1 mM) before use.

Nanoparticle (np-) $Al_2O_3$ was prepared by adding 100 µl of $Al_2O_3$ np solution to 15 ml of anhydrous isopropyl alcohol to give a 1:150 vol ratio solution, which was stirred overnight before use. The perovskite solution was prepared first as a 1.4 M stock (typically 5 ml) containing (per ml) 199.8 mg FAI, 61.8 mg CsI, 403.4 mg $PbI_2$ and 192.7 mg $PbBr_2$ in 4:1 DMF:DMSO, which was stirred overnight. Before use, this solution was diluted in a 6:1 volume ratio with 4:1 DMF:DMSO and agitated plenty. Similarly, a 10 mol% EDA stock was prepared, containing 1.2 ml perovskite, 200 µl 4:1 DMF:DMSO and 11.2 µl EDA. The 0.5-3 mol% EDA additive solutions were then prepared using volumetric mixtures of these two stock solutions, with plenty of mixing and used ~1-2 hours after the EDA was added. PCBM solution was 20 mg $ml^{-1}$ in a 3:1 chlorobenzene:dichlorobenzene mixture, mixed plenty and then filtered with 0.45 µm PTFE filter before use. BCP solution (0.5 mg $ml^{-1}$) in IPA was stirred overnight and heated to 70 °C if required and filtered before use.

Pre-patterned indium tin oxide substrates were cleaned by scrubbing with 1% Decon90 solution, before sonicating sequentially in 1% Decon90, ionized water, acetone and isopropyl alcohol for for ~10 minutes each, with rinsing between each step. Substrates were dried under $N_2$ and UV ozone treated for 30 minutes directly before use and transferred to an $N_2$ glovebox for fabrication with constant purging. Me-4PACz SAM solution (150 µl, 0.33 $mgml^{-1}$ in ethanol) was spread over the substrate, and after 10 seconds spin coated at 3000 rpm for 30 s, before annealing at 100 °C for 10 minutes. After cooling, 70 µl of np-$Al_2O_3$ was dynamically spin-coated at 6000 rpm for 30 s, with ~ 1 min annealing at 100 °C. The perovskite was spin coated at 1000 rpm (200 $rpms^{-1}$ ramp) for 10 s, followed by 35 s at 5000 rpm (1000 $rpms^{-1}$ ramp), with 170 µl of the solution dynamically coated 3 seconds into the program, and 330 µl of anisole was dropped as an antisolvent 10 s before the end of the program. With increasing EDA content, the intermediate film was observed to be highly smooth. Films were transferred to the hotplate after a few seconds and annealed at 100 °C for 45 minutes. After cooling, 50 µl of PCBM was dynamically coated at 2000 rpm onto the perovskite surface, and dried at 100 °C for 5-10 minutes. Once cool, BCP was dynamically coated at 5000 rpm and dried at 100 °C for 2 minutes. After removing to air and patterning the devices, 100 nm of Au was thermally evaporated under high vacuum at a rate of $0.1 - 1.2$ $\text{Ås}^{-1}$. Devices were left overnight in $N_2$ before testing.

*Device characterization:*

Current-voltage (J–V) and maximum power point (MPP) measurements were measured using a Keithley 2400 source meter (Keithley Instruments) in ambient air with 100 mWcm$^{-2}$ AM 1.5g irradiance from a Wavelabs SINUS-220 simulator, and in the dark. The active area of each solar cell was defined using a black metal aperture mask to either 0.25 or 1.00 cm$^2$ within a black holder. Reverse J–V scans were measured from 1.3 V (forward bias) to beyond short-circuit (-0.1 V) and forward scans were from short-circuit to forward bias, both at a scan rate of ~0.24 Vs$^{-1}$. MPP tracking measurements were performed using a "gradient descent" algorithm for 30 s to obtain the steady-state power conversion efficiency. The intensity of the solar simulator is automatically calibrated and was additionally verified periodically using a KG5-filtered Si reference photodiode (Fraunhofer ISE).

*NMR:*

A two-channel Bruker Avance III HD Nanobay 400 MHz instrument running TOPSPIN 3 equipped with a 5 mm z-gradient broadband/fluorine observation probe was used, with the signal from residual non-deuterated DMSO solvent used to reference the spectra.

*GIWAXS:*

*In situ* grazing-incidence wide-angle X-ray scattering data was acquired at the I07 undulator beamline at Diamond Light Source. Solutions were deposited using an *in situ* blade coater contained in an $N_2$ environment incorporating a syringe driver, coating surface, motorised blade, integrated hotplate and an $N_2$ outlet directed at the sample acting as an air knife. Prior to data acquisition, solutions were deposited onto cleaned glass substrates, and coated with a blade with a shim height of 100 um and coating speed of 9 mms$^{-1}$, with gas quenching applied continuously from 30 s. Monochromatic X-rays with energy 10 keV were incident on the sample at a grazing angle of 1°, with scattering collected by a Pilatus 2M (DECTRIS) hybrid photon-counting detector at a distance of 365 mm, calibrated using an LaB6 standard. 2D detector images were acquired every 0.2 s. Data reduction was performed using scripts based on the pyFAI and pygix libraries.[5]

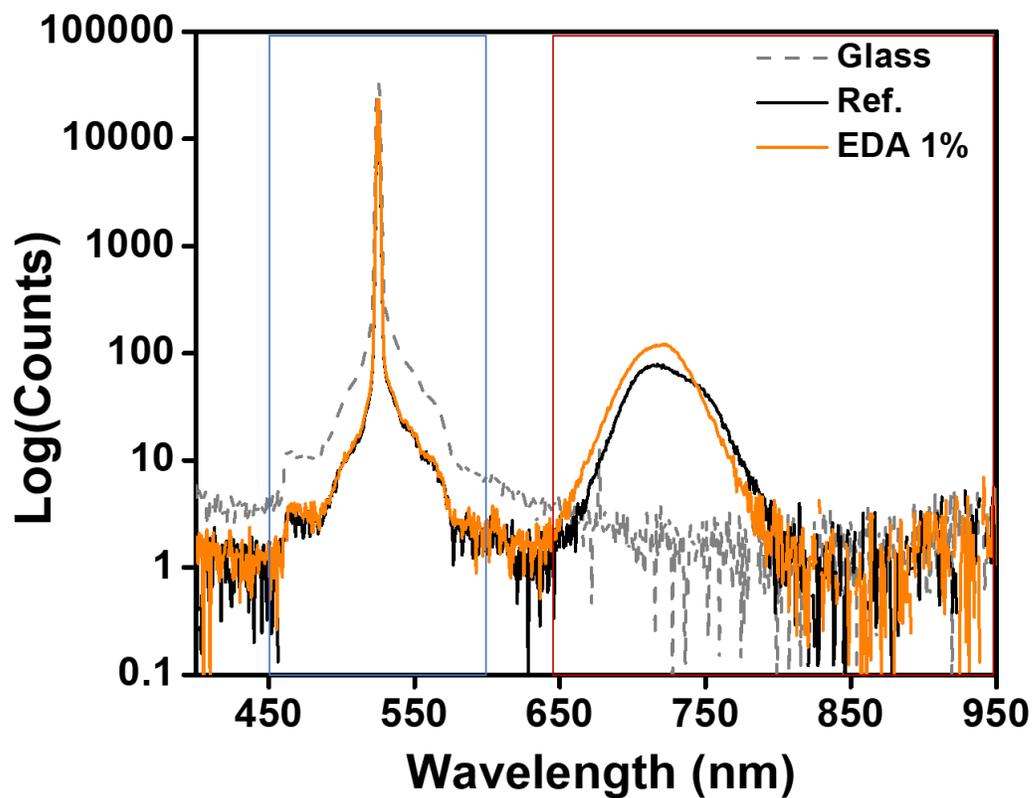

*Figure S1*: PL spectra from integrating sphere coupled with a 532 nm laser at 60 mA/cm$^2$. Blue lines show the excitation region and the red lines the emission region. The dashed grey spectrum is the one of a bare glass film used as blank.

|  | A1 | $\tau_c$ | $\beta$ | $\tau_{stretch}$ |
|---|---|---|---|---|
| **EDA-0** | 1.21 | 0.58 | 0.17 | 411 |
| **EDA-1** | 1.05 | 301 | 0.49 | 627 |

*Table S1*: Results from the stretched exponential fitting function of photoluminescence decays shown in Figure 1b.

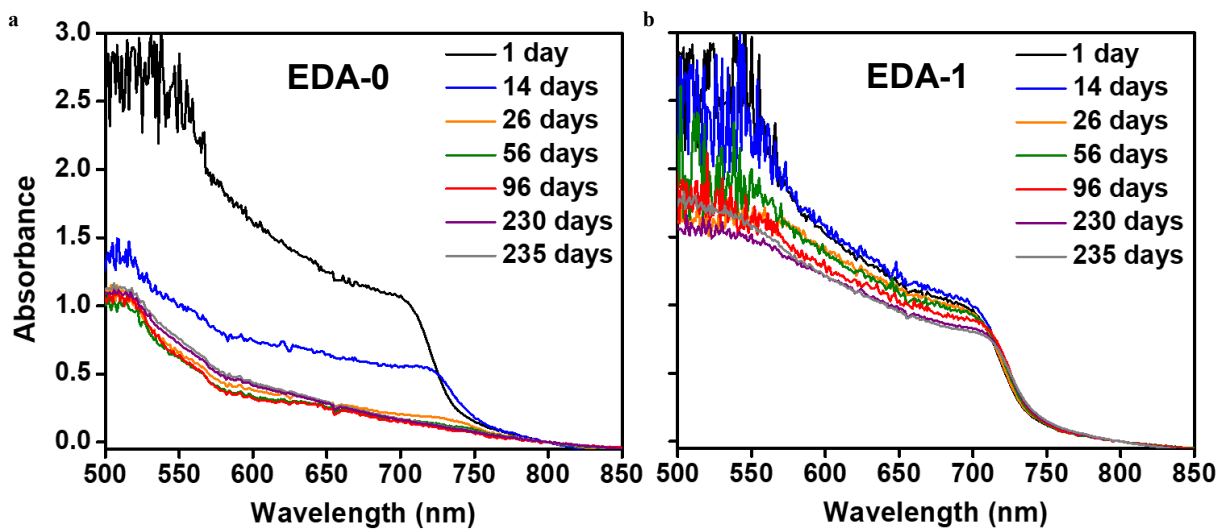

*Figure S2:* *UV-Vis of EDA-0., EDA-1 over 235 days of aging. The absorbance was corrected for scattering by subtracting the background signal from 850 to 1200 nm. Samples were stored unencapsulated in the dark at an average room temperature of 19.5 °C and 31% RH.*

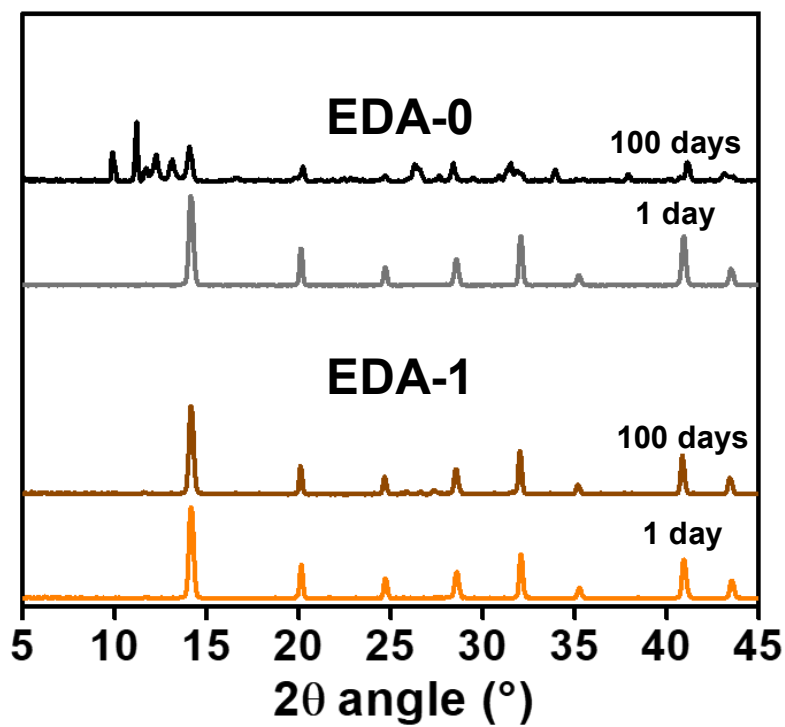

*Figure S3:* *XRD patterns of EDA-0 and EDA-1 after 100 days of aging. Samples were stored unencapsulated in the dark at an average room temperature of 19.5 °C and 31% RH.*

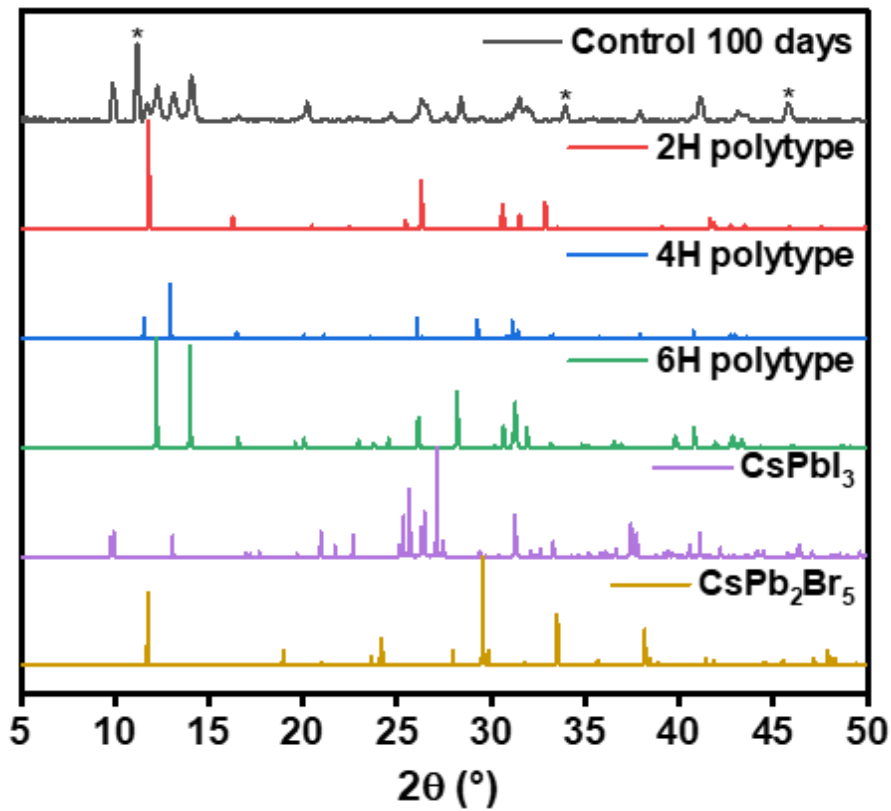

**Figure S4:** *XRD pattern of Control film after 100 days of aging compared against simulated powder XRD patterns for likely decomposition products 2H, 4H and 6H FAPbX$_3$ polytype phases,[6] CsPbI$_3$ (PDF card 01-084-2969) and CsPb$_2$Br$_5$.[7] Additional starred reflections not accounted for by these phases are consistent with a moisture-induced phase reported by Hu et al as CsPb$_2$I$_4$Br.[8] Further differences in expected peak positions may be explained by halide mixtures in the above phases.*

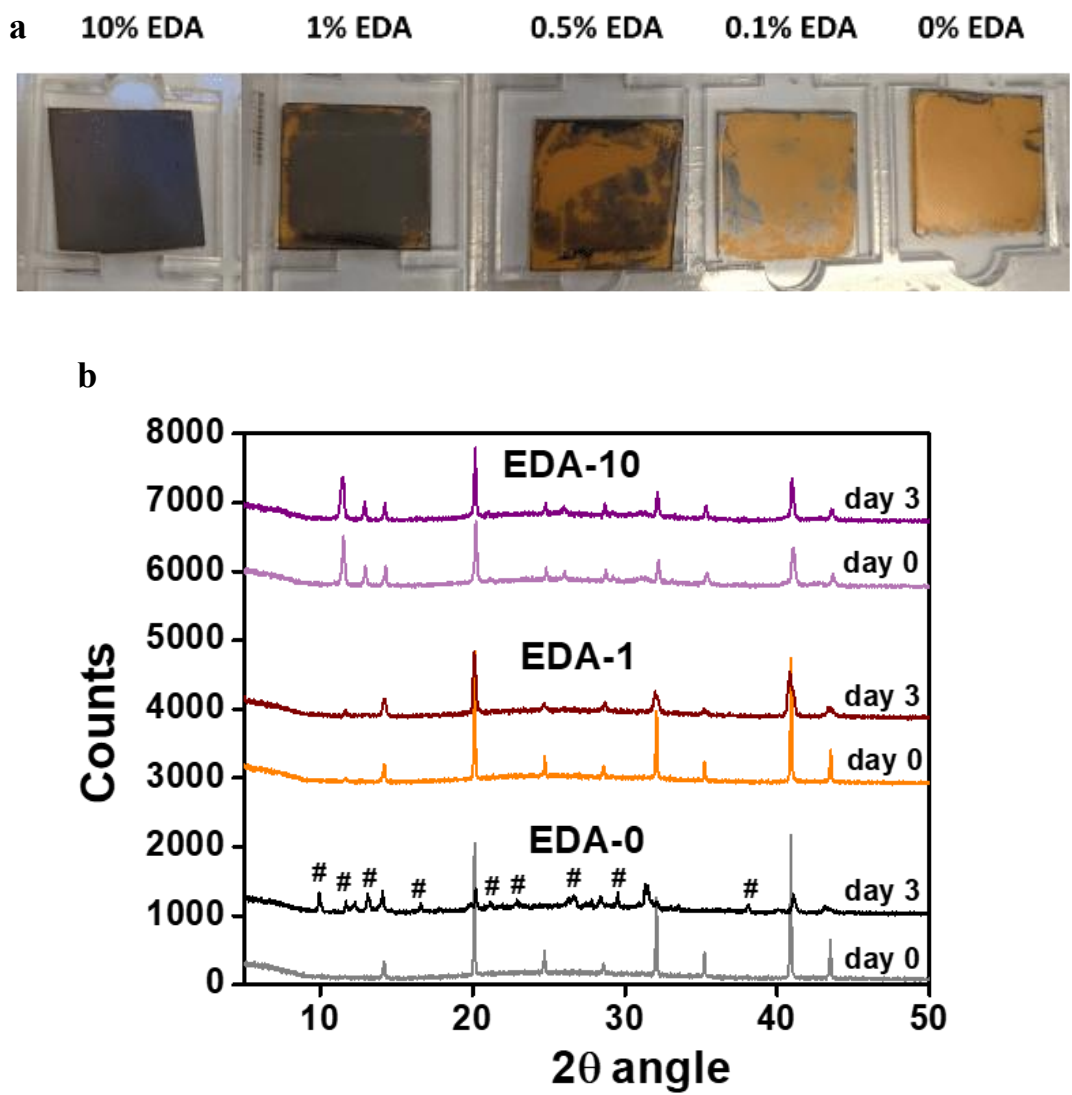

*Figure S5: (a) Pictures of $FA_{0.83}Cs_{0.17}Pb(I_{0.75}Br_{0.25})_3$ films made at the University of Oxford after 3 days aging in ambient air, ambient room lighting, and uncovered (air flow). (b) XRD on films shown in part a. Grey, orange and pink patterns are from the films as prepared of EDA-0, EDA-1 and EDA-10. Black, dark red, and purple patterns are from the films after 3 days aging in air. Peaks marked with \* are associated with 2H and 4H polytypes, please see discussion in SI Note 1. Peaks marked # denote decomposition products patterns from EDA-0 (see Fig. S4).*

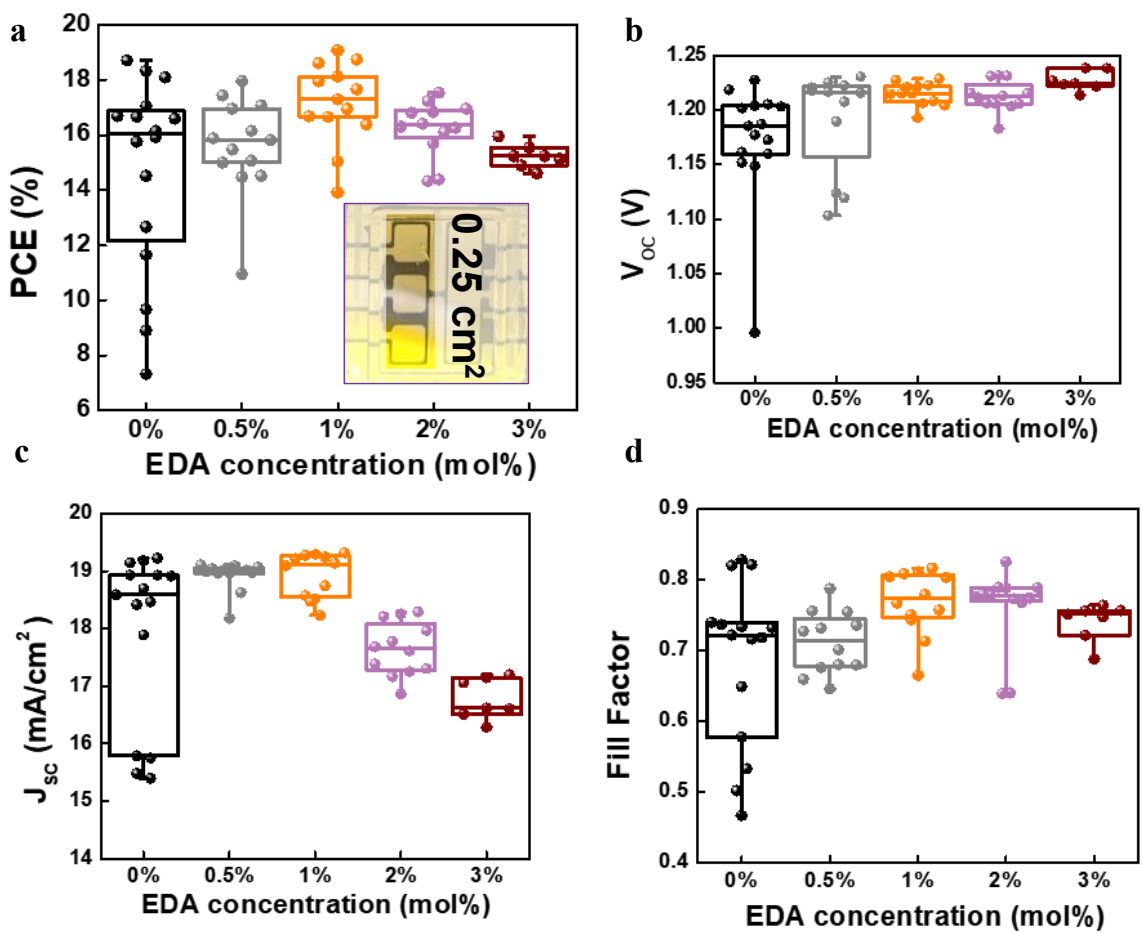

*Figure S6*: *(a) Reverse sweep power conversion efficiency (PCE) (b) Open circuit voltage ($V_{OC}$) (c) Short circuit current ($J_{SC}$) and (d) fill factor (FF) of devices with 0 (black), 0.5 (grey), 1 (orange), 2 (violet) and 3 (brown) mol% of EDA in the perovskite layer. Inset: Image showing area of the device pixels.*

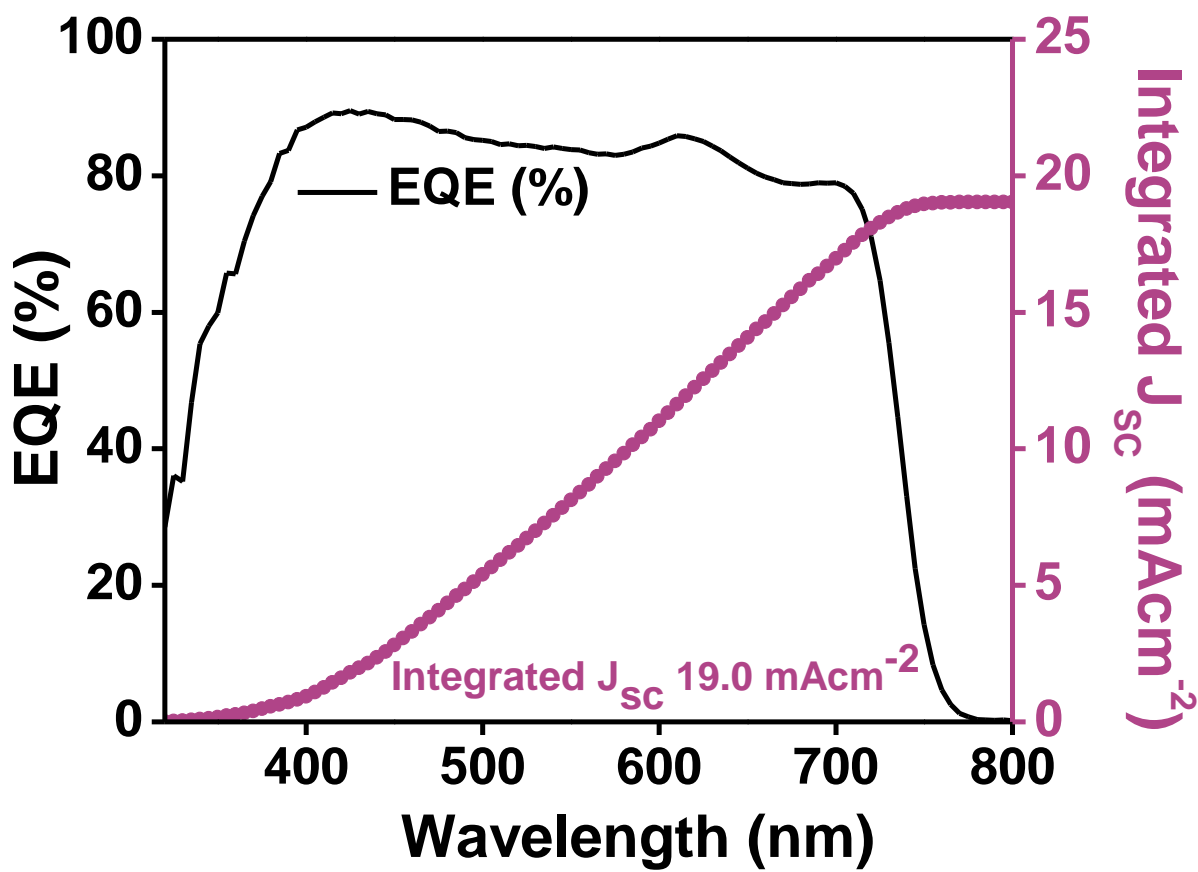

*Figure S7:* *Champion device (as shown in Figure 2b) external quantum efficiency (EQE) spectrum (black) and corresponding short-circuit current density ($J_{SC}$, purple dotted line) calculated by integration of each EQE spectrum with the global AM1.5G solar spectrum.*

| DOI | Authors | Year | Bandgap (eV) | $V_{OC}$ (V) | PCE (%) |
|---|---|---|---|---|---|
| 10.1038/nenergy.2017.135 | Wang et al. | 2017 | 1.61 | 1.14 | 20.6 |
| 10.1021/acs.chemmater.8b02970 | Liu et al. | 2018 | 1.62 | 1.08 | 19.02 |
| 10.1039/c8ta04936j | Liu et al. | 2018 | 1.625 | 1.12 | 17.65 |
| 10.1039/c5ee03874j | Saliba et al. | 2016 | 1.63 | 1.058 | 15.4 |
| 10.1039/c8ta05795h | Svanström et al. | 2018 | 1.64 | 1.02 | 15 |
| 10.1021/acsenergylett.8b01165 | Werner et al. | 2018 | 1.65 | 1.082 | 15.2 |
| 10.1126/science.aba1631 | Lin et al. | 2020 | 1.66 | 1.16 | 17.3 |
| 10.1002/adma.201801562 | Long et al. | 2018 | 1.67 | 1.19 | 17.92 |

| DOI | Author | Year | | Eg (eV) | Voc (V) | PCE (%) |
|---|---|---|---|---|---|---|
| **10.1002/aenm.201902353** | Raiford et al. | 2019 | | 1.68 | 1.162 | 18.43 |
| **10.1021/acsenergylett.8b01165** | Werner et al. | 2018 | | 1.69 | 1.107 | 14.8 |
| **10.1021/acsami.9b17241** | Bett et al. | 2019 | | 1.7 | 1.166 | 14.6 |
| **10.1016/j.mtener.2017.10.001** | Yang et al. | 2017 | | 1.715 | 1.085 | 14.9 |
| **10.1002/aenm.201701048** | Zhou et al. | 2017 | | 1.72 | 1.24 | 18.13 |
| **10.1021/acsenergylett.8b02179** | Jaysankar et al. | 2018 | | 1.72 | 1.22 | 13.8 |
| **10.1016/j.joule.2019.05.009** | Palmstrom et al. | 2019 | | 1.7 | 1.2 | 19.2 |
| **10.1021/acsenergylett.7b01255** | Bush et al. | 2018 | | 1.68 | 1.1 | 15.4 |
| **our result** | Taddei et al. | 2022 | | 1.685 | 1.22 | 19.07 |

*Table S2*: *List of references from which we extracted the PCE and $V_{OC}$ values reported in Figure 3c and 3d.*

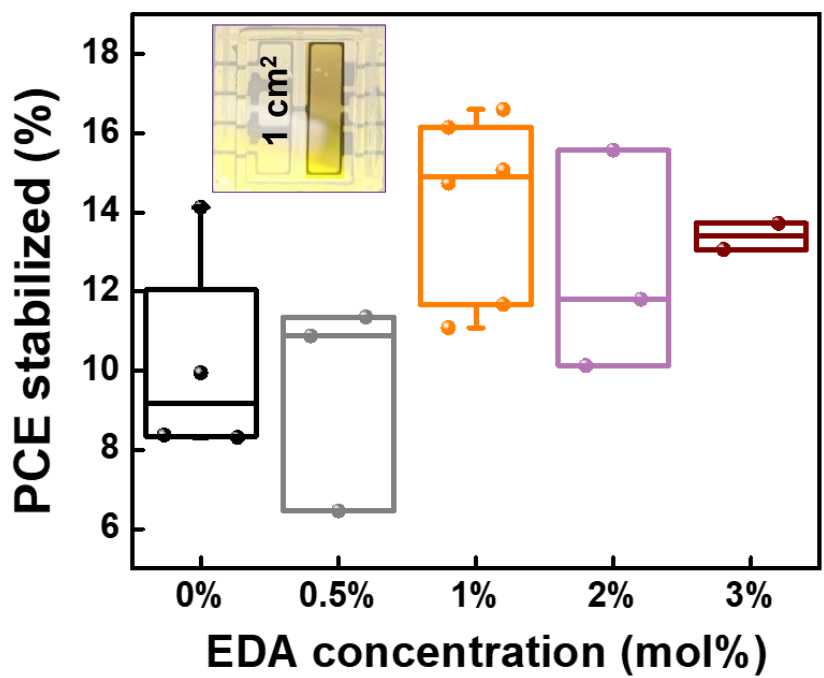

*Figure S8*: *Maximum power point extracted efficiency of 1 $cm^2$ area devices with 0 (black), 0.1 (light grey), 0.5 (grey), 1 (orange), 2 (violet) and 3 (brown) mol% of EDA in the perovskite layer. Inset: Area of device pixel.*

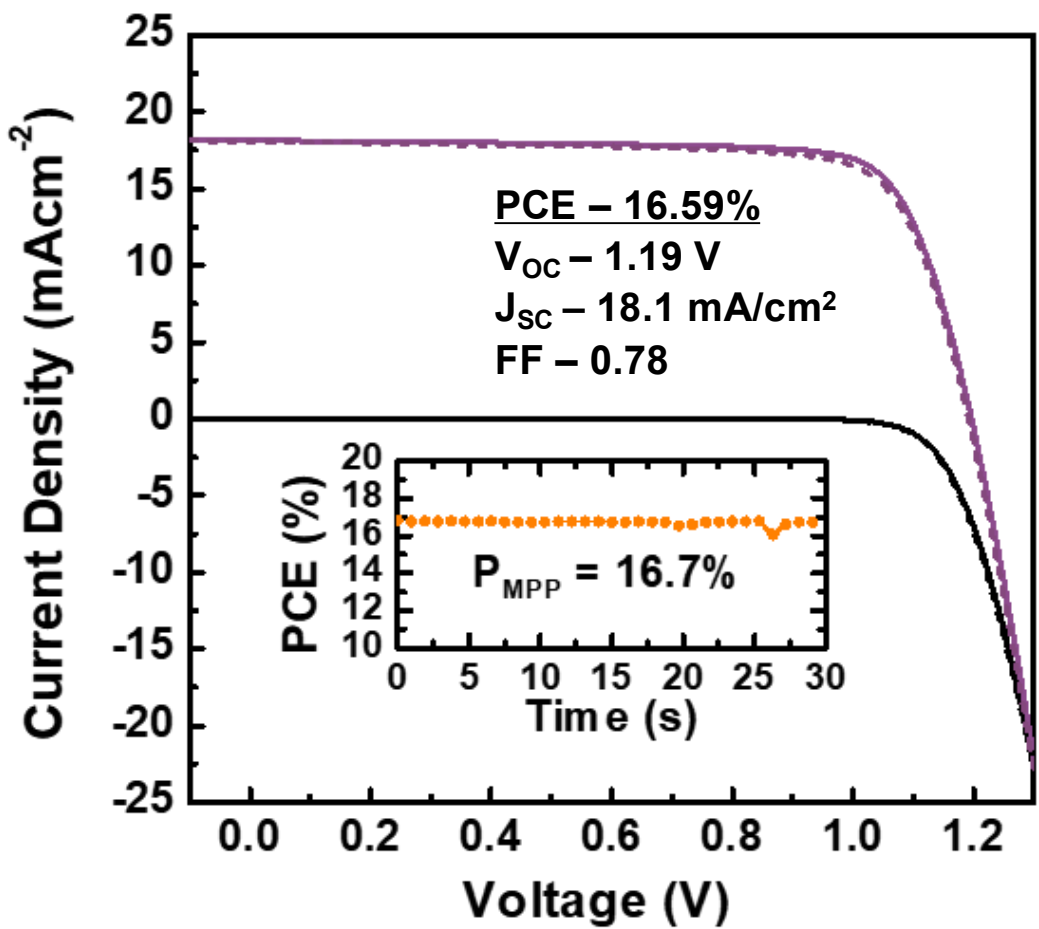

*Figure S9*: *JV curves of the EDA-1 large area champion device under 1 sun simulated solar illumination, recorded in forward (from short to open circuit) and reverse (from open to short circuit) bias. Inset shows the MPP-tracked device efficiency.*

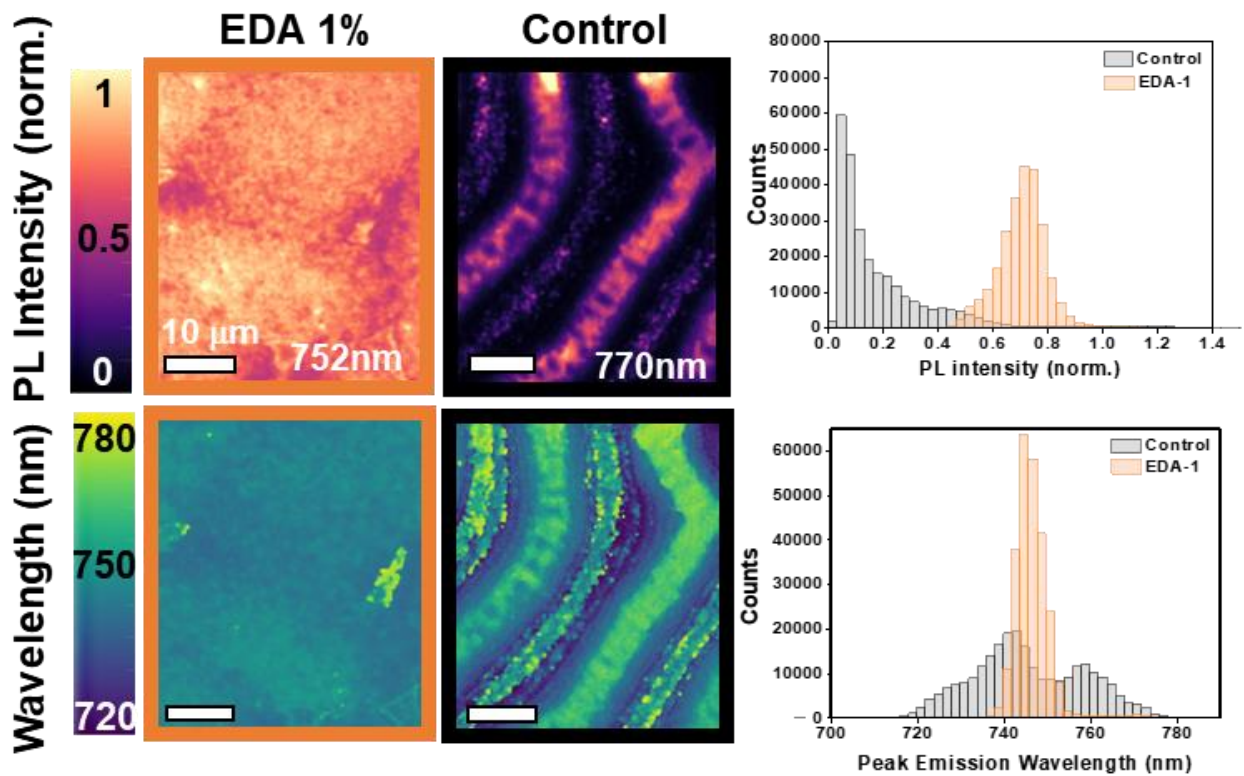

*Figure S10:* *(a) Hyperspectral PL intensity map of the Control film (black box) and EDA-1 (orange box). PL images are shown at the maximum emission wavelength ($\lambda_{max}$) for each sample as stated on the figure (Control $\lambda_{max}$ = 772 nm, EDA-1 $\lambda_{max}$ = 752 nm). Excitation was achieved with a white lamp with short pass filter at 500 nm. The emission was recorded using a 500 nm dichroic and 650 nm long pass filter. (b) Histogram of the PL intensity distribution at the maximum emission wavelength for the Control (grey) and EDA-1 (orange) samples. (c) Peak emission wavelength map of the Control and EDA-1 films (d) Histogram of the distribution of peak emission wavelengths for the Control (grey) and EDA-1 (orange) samples.*

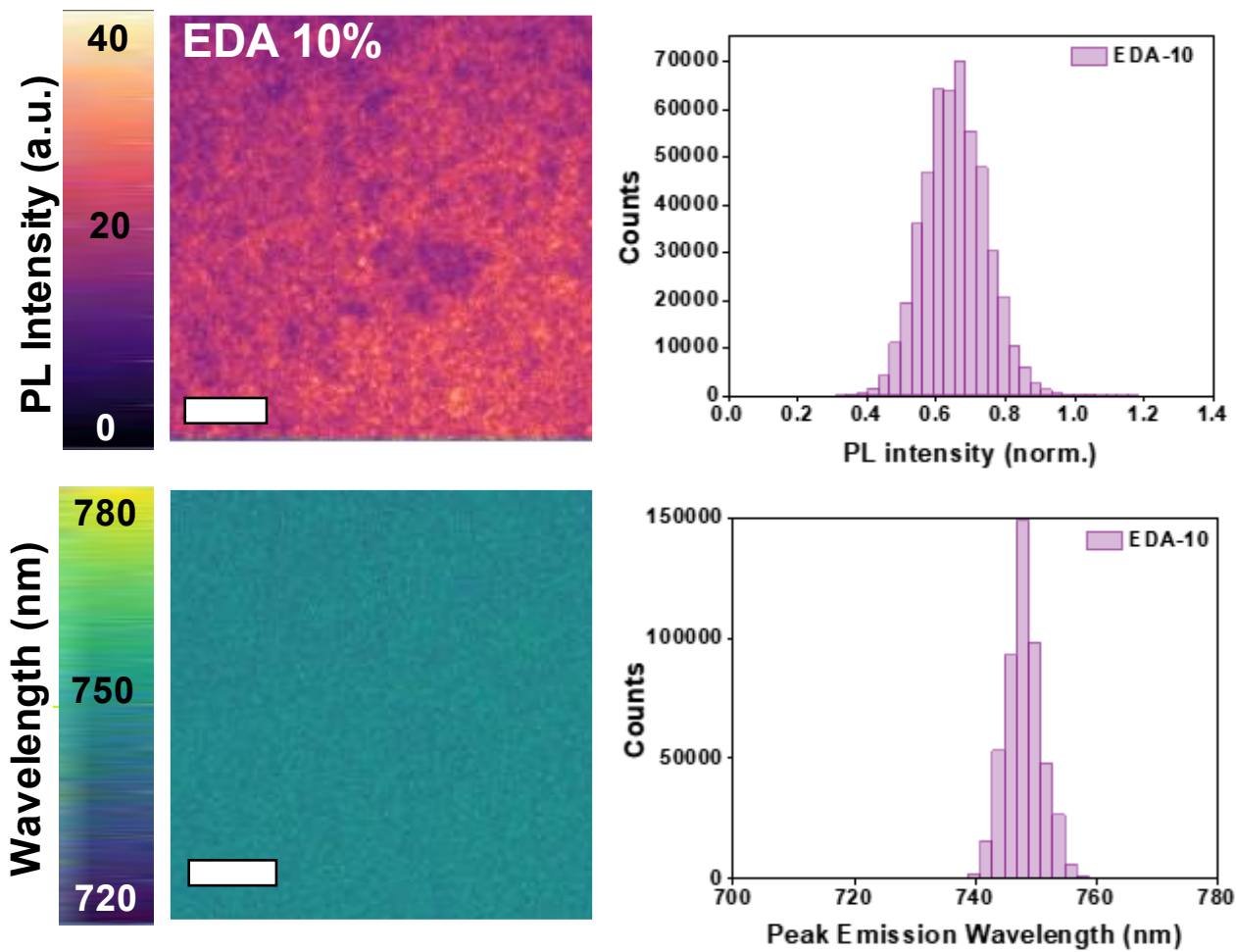

*Figure S11:* (a) Hyperspectral PL intensity map of the EDA-10 sample. Excitation was achieved with a white lamp with short pass filter at 500 nm. The emission was recorded using a 500 nm dichroic and 650 nm long pass filter. (b) Histogram of the PL intensity distribution at the maximum emission wavelength. (c) Peak emission wavelength map. (d) Histogram of the distribution of peak emission wavelengths.

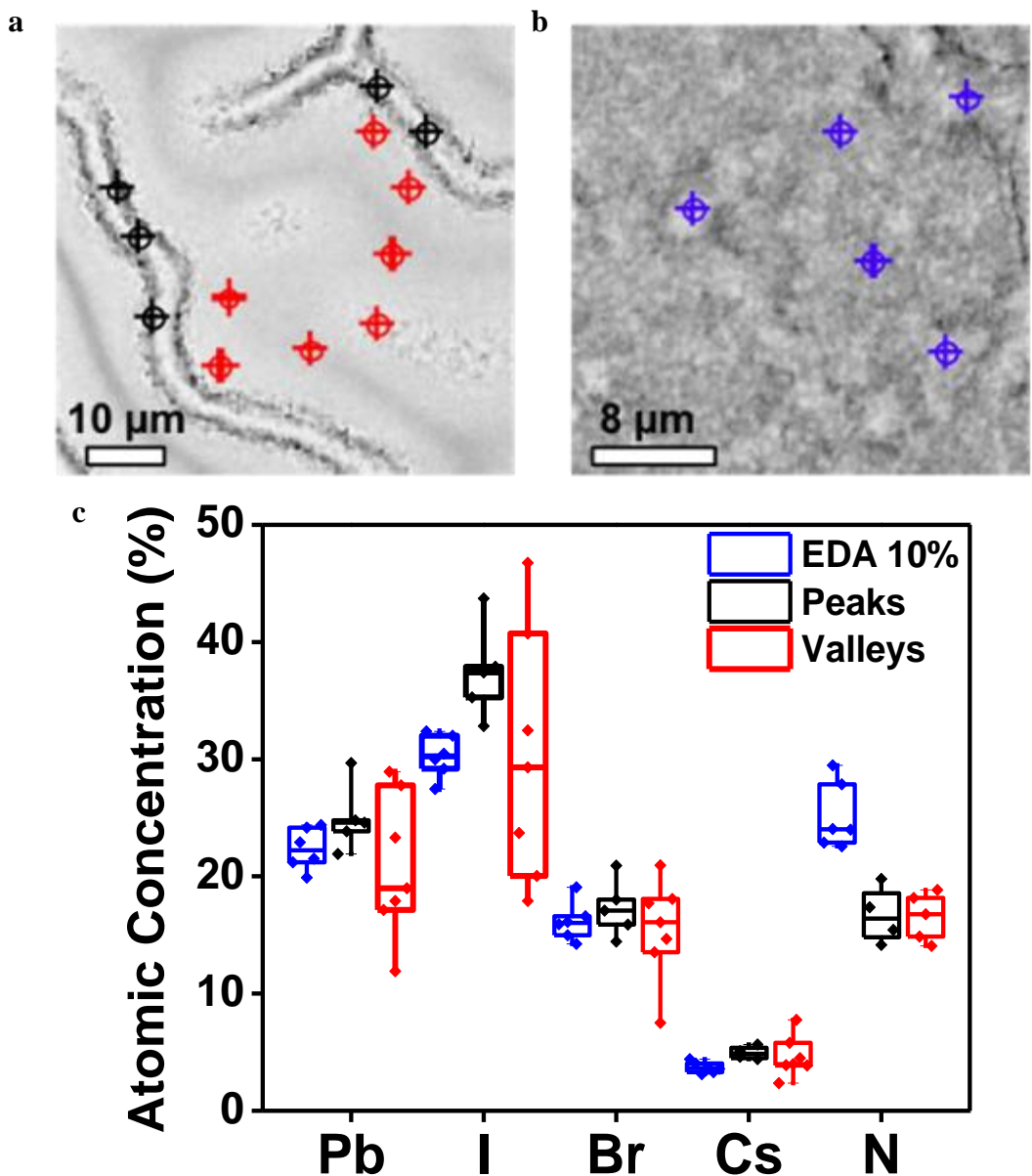

*Figure S12*: *Scanning electron microscopy image of Reference (a) and EDA-10 (b). The electron gun power was 15 kV from which we extracted elemental statistics by EDX shown in (c). The atomic concentration was analyzed in multiple points at the "peaks" of the wrinkles (black), in the "valleys" (red) and at several locations across the EDA-10 sample (blue).*

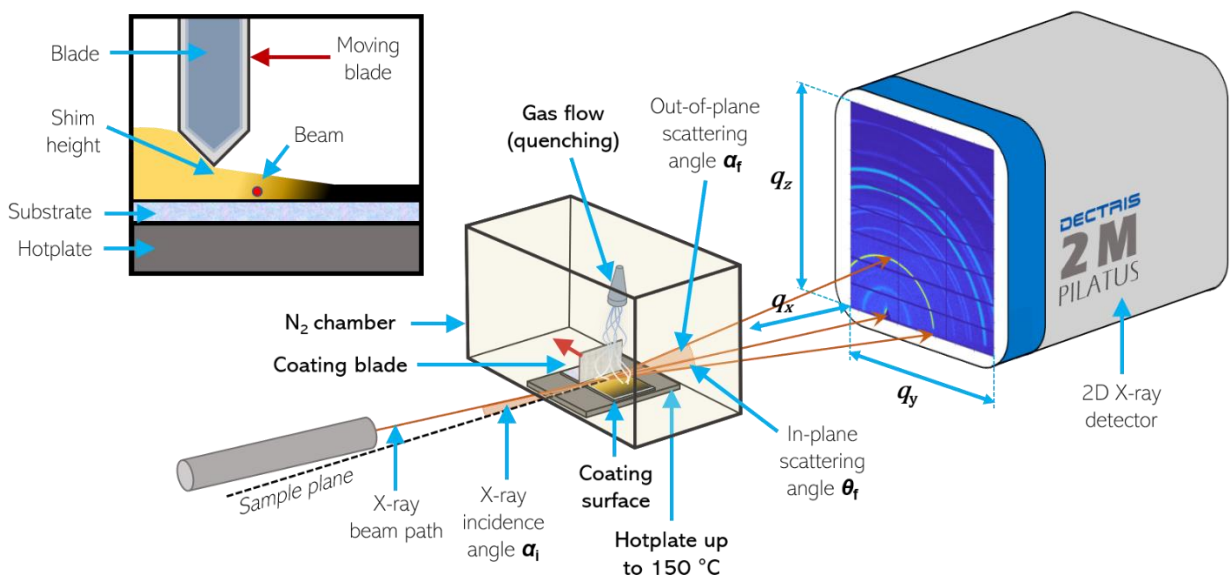

*Figure S13:* Schematic illustration of GIWAXS experiments combining in-situ blade coater (moving orthogonal to the synchrotron beam), $N_2$ gas quenching to remove solvent and integrated hotplate for annealing. Scattering is collected with a 2D hybrid photon counting detector. Further details are given in the experimental methods.

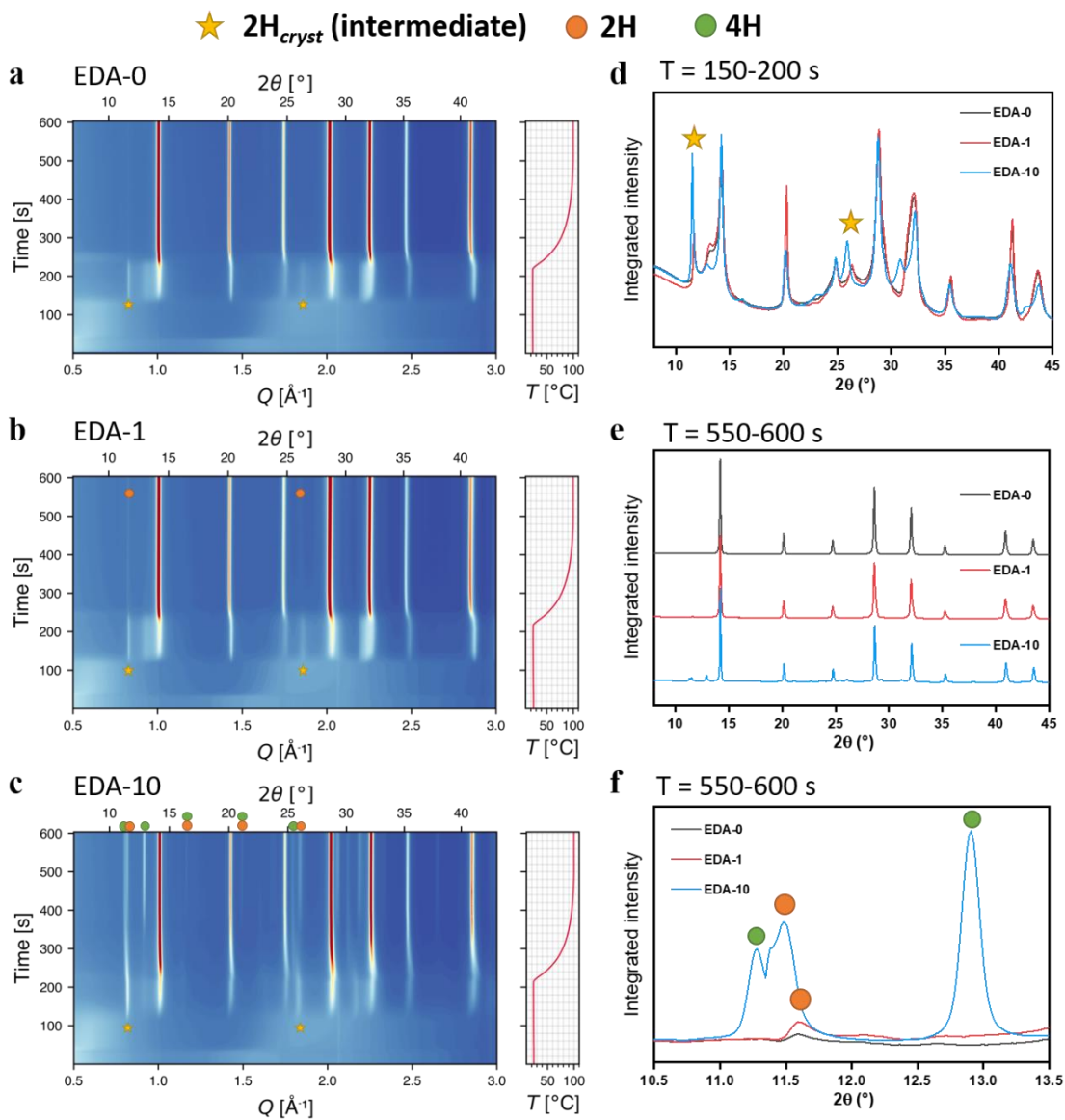

*Figure S14:* In situ GIWAXS (azimuthally integrated from acquired 2D images) of blade coated solutions of (a) EDA-0, (b) EDA-1 and (c) EDA-10 samples. Peaks indicated with a yellow star indicate the characteristic reflections from a 2H polytype phase, forming as an intermediate during crystallisation, and orange and green circles indicate 2H and 4H polytype phases present after annealing. (d) Integrated 1D intensity plots from all frames between 150 and 200 s into the measurement, highlighting the initially formed 2H intermediate, which is stronger for EDA-10. (e) Integrated 1D intensity plots after heating to 100 ºC (all frames from 500-550 s), showing the dominant phase in all cases is a pseudo-cubic perovskite. (f) Highlight of part (e) showing weak scattering from a polytype at $2\theta = 11.6º$ for the EDA-1 film, and 2H and 4H polytypes present for EDA-10.

**SI Note 1 – Phase segregation, halide sequestration and bandgap widening with EDA addition.**

To understand the effect of EDA on phase behavior in thin films, we acquired XRD from films with higher additive concentrations (up to 40 mol%, EDA-40), with full patterns shown in **Figure S15**. In **Figure S16** we follow the position of the (100) reflection from the α and polytype phases. The α-phase shifts to larger 2θ angles (smaller lattice parameter) from EDA-0 to EDA-40, commensurate with the 2H/4H polytype mixture tending towards entirely 2H (reduced corner-sharing). In **Figure S17** and **S18** we show both the perovskite band-edge and the PL progressively increase in energy as the EDA concentration is increased over the same range. These changes in the XRD, PL and absorbance are all consistent with formation of an iodide-rich secondary phase, resulting in a bromide- and cesium-rich α-phase with wider bandgap. Specifically, at high concentrations, with enough of a proposed large cation present, this cation forces segregation of a pure-iodide 2H phase (**Figure S19**) while at lower concentrations (EDA-10) a 4H polytype is formed (most likely containing some bromide), having a less significant effect on altering the 3C perovskite phase composition.

To confirm the proposed perovskite bandgap widening mechanism, arising from the formation of a 1D imidazolinium lead iodide secondary phase and Cs- and Br-rich α-phase, we followed the same additive approach and investigated films of two Br-free FA-rich compositions, $FA_{0.95}Cs_{0.05}PbI_3$ and $FA_{0.83}Cs_{0.17}PbI_3$. As expected, Tauc plots from UV-Vis absorption measurements (**Figure S20**) showed only a very slight widening of the optical bandgap from the Control film to EDA-40, from 1.57 eV to 1.6 eV for $FA_{0.83}Cs_{0.17}PbI_3$ and 1.54 eV to 1.55 eV for $FA_{0.95}Cs_{0.05}PbI_3$. The bandgap from these for EDA0-40 are summarized in **Figure S21**. Moreover, in **Figure S22** we verified that the effect of bandgap widening with EDA is happening with different bromide ratios across the $FA_{0.83}Cs_{0.17}Pb(I_{1-x}Br_x)_3$ compositional space (with x= 0.15-0.28) after addition of 10 mol% of EDA. The bandgap trend is plotted against bromide content in **Figure S23.**

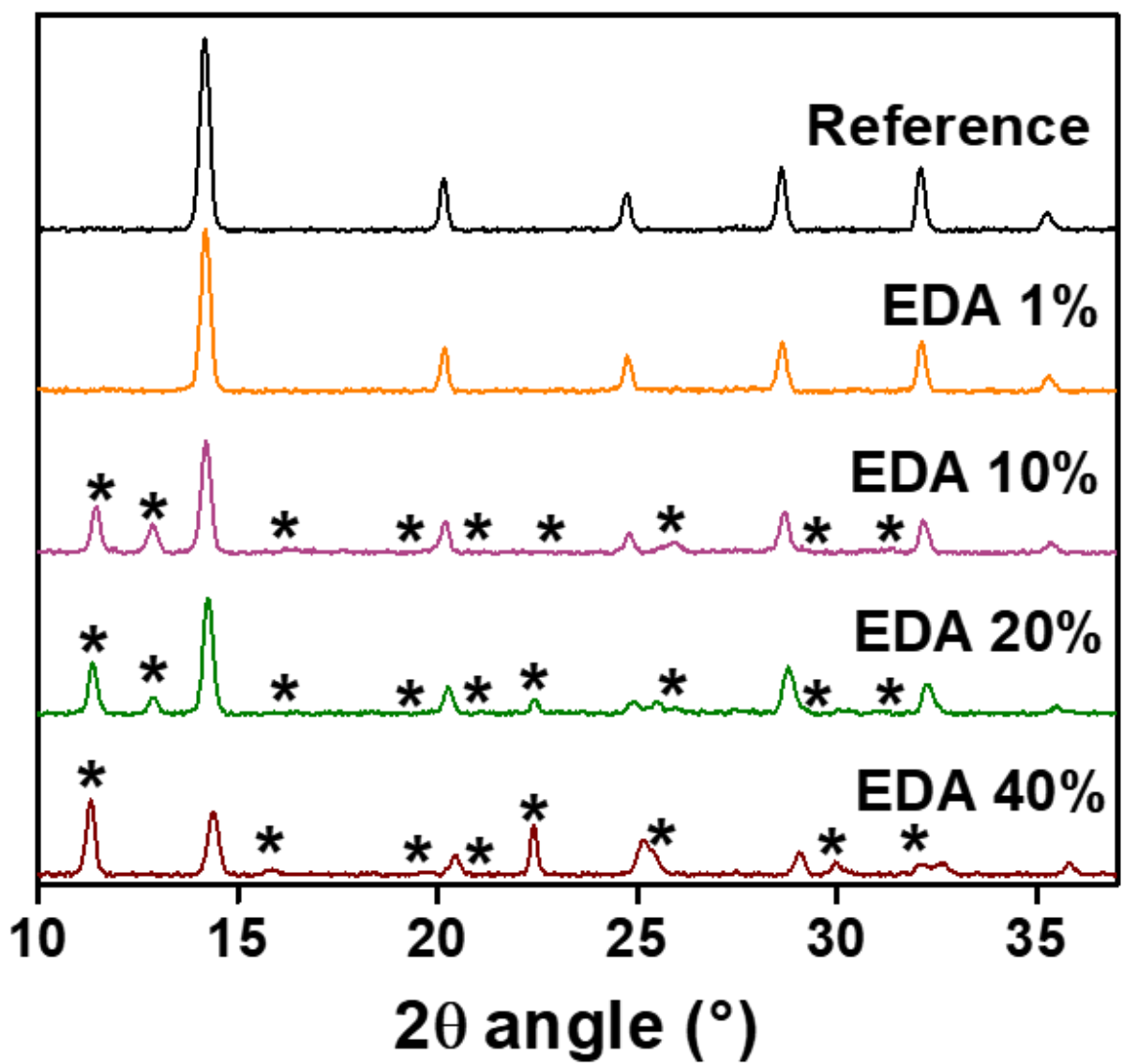

*Figure S15:* XRD pattern of EDA-0 (Control, black), EDA-1 (EDA 1%, orange), EDA-10 (EDA 10%, purple), EDA-20 (EDA 20%, green), EDA-40 (EDA 40%, brown) peaks indicated with * indicates the characteristic reflections from 2H and 4H polytype phases (simulated diffraction patterns are given in Figure S4)

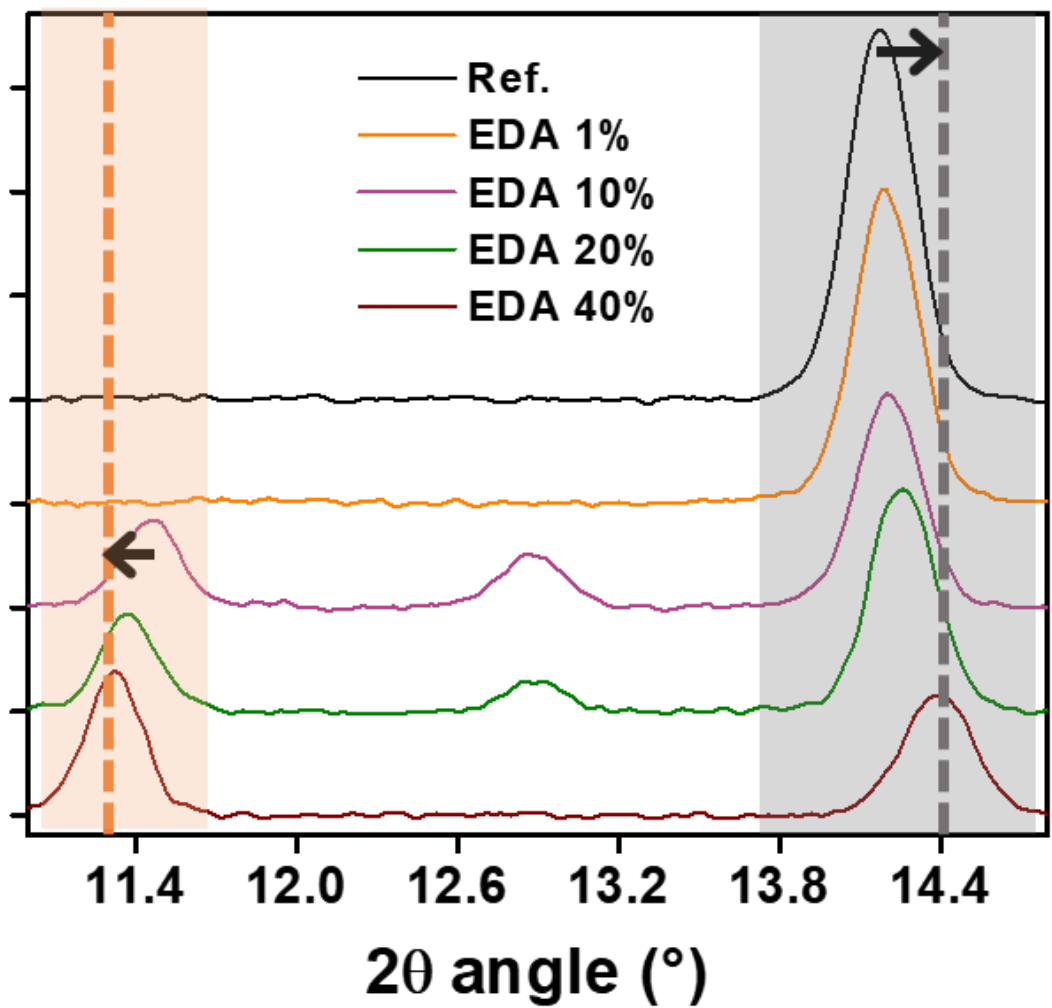

*Figure S16:* Zoom on 2θ angles 11.2° to 14.6°. 2H (100) / 4H (100) peaks are highlighted in orange and cubic (100) in grey. Arrows indicate a shift of the 2H/4H (100) and cubic (100) phase with increasing EDA. Reduction of the peak at ~12.9° from the 4H polytype confirms the phase mixture in the film shifting to entirely 2H polytype at the highest EDA concentration.

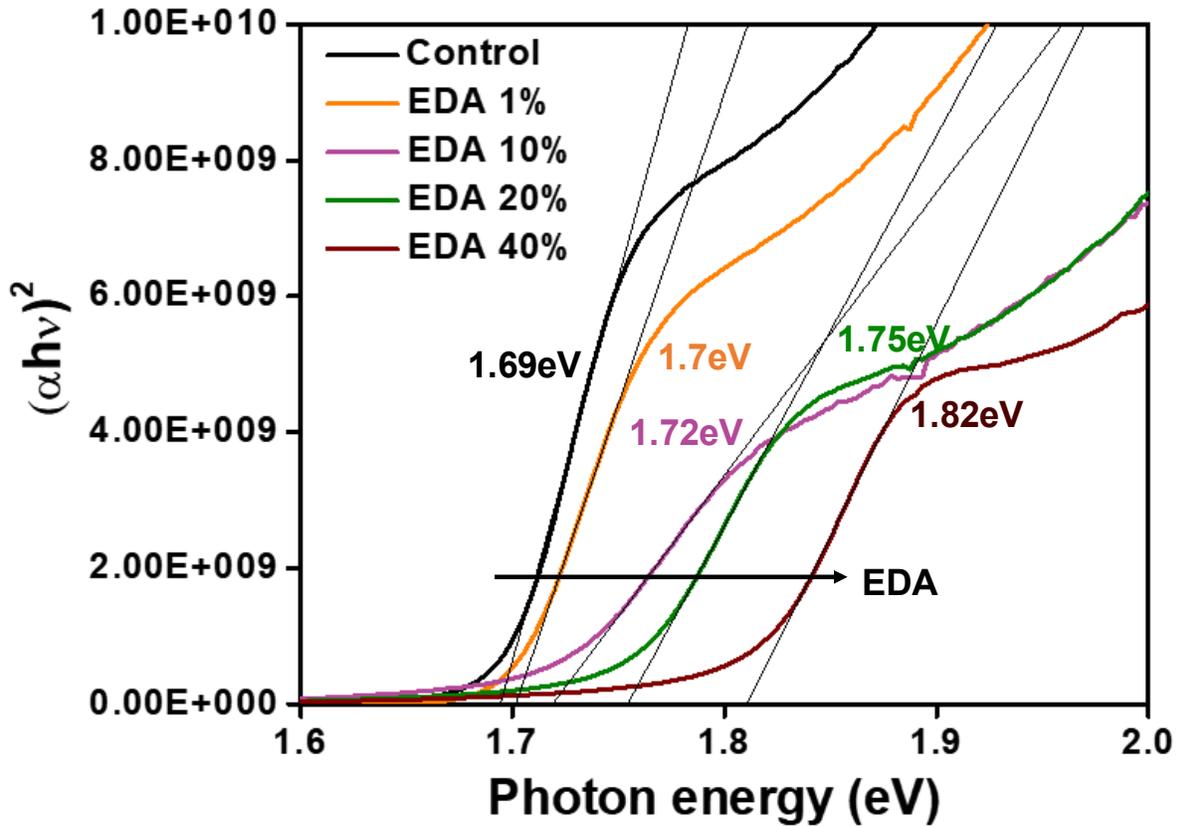

*Figure S17*: Tauc plot extracted from UV-Vis of EDA-0 (Control, black), EDA-1 (EDA 1%, orange), EDA-10 (EDA 10%, purple), EDA-20 (EDA 20%, green), EDA-40 (EDA 40%, brown). Arrows show higher $E_g$ with more EDA added.

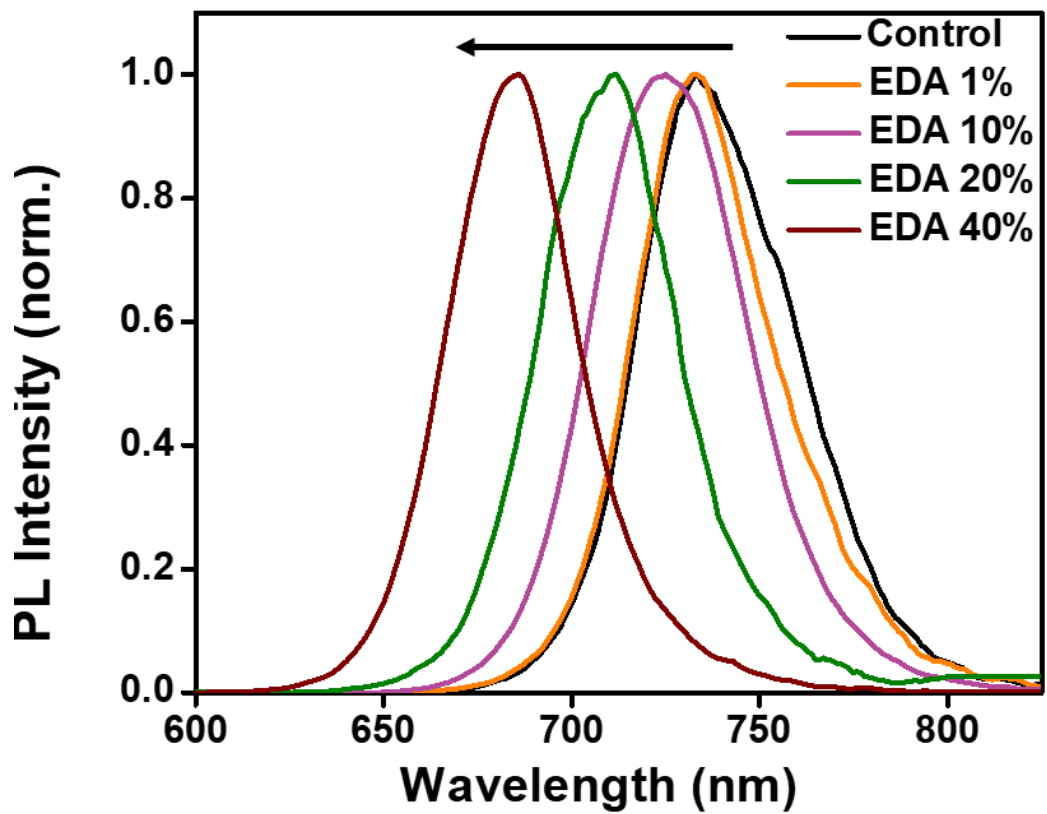

*Figure S18:* *Normalized steady state PL of EDA-0 (Control, black), EDA-1 (EDA 1%, orange), EDA-10 (EDA 10%, purple), EDA-20 (EDA 20%, green), EDA-40 (EDA 40%, brown) at $\lambda_{exc}$ = 500nm.*

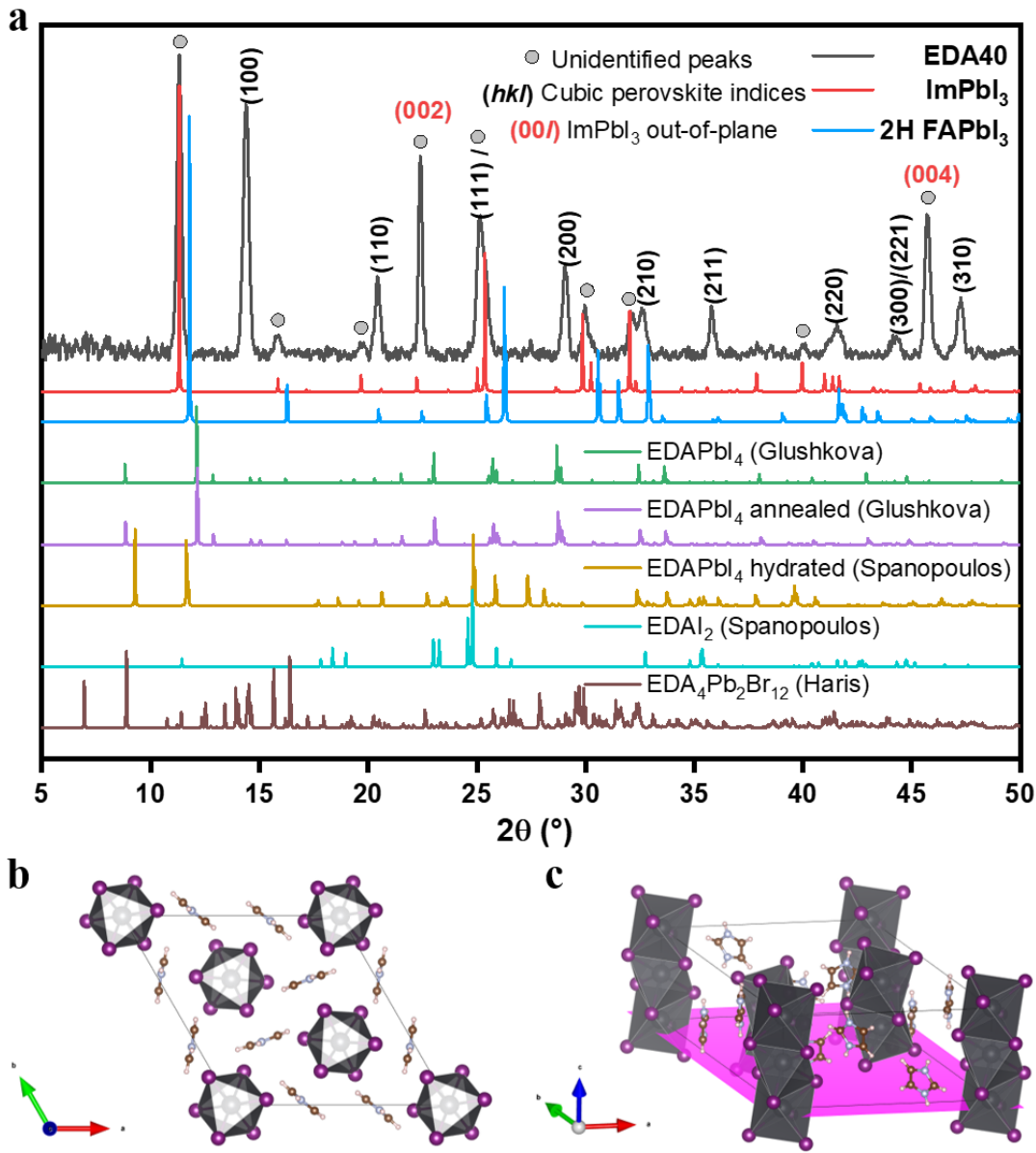

***Figure S19:*** *(a) XRD pattern of sample with EDA 40 mol% addition (grey), simulated imidazolium lead iodide (ImPbI₃) 2H phase pattern (red) and FAPbI₃ 2H polytype (blue). Below are further simulated XRD patterns for reported EDAPbI₄, EDAI₂ and EDA₄Pb₂Br₁₂ phases, showing these cannot account for the additional reflections.[9–11] Parts (b) and (c) are projections of the ImPbI₃ 2H phase simulated using VESTA. In this phase the 1D iodoplumbate chains are distorted from the idealized hexagonal P6₃/mmc unit cell for δ-FAPbI₃ due to steric effects arising from the flat imidazolium cation and are instead indexed as a larger hexagonal P6₃/m unit cell. The pink plane in part c) denotes the (00l) planes which are higher intensity reflections (marked red) in the EDA-40 XRD, confirming the iodoplumbate chains are oriented in the out-of-plane direction.*

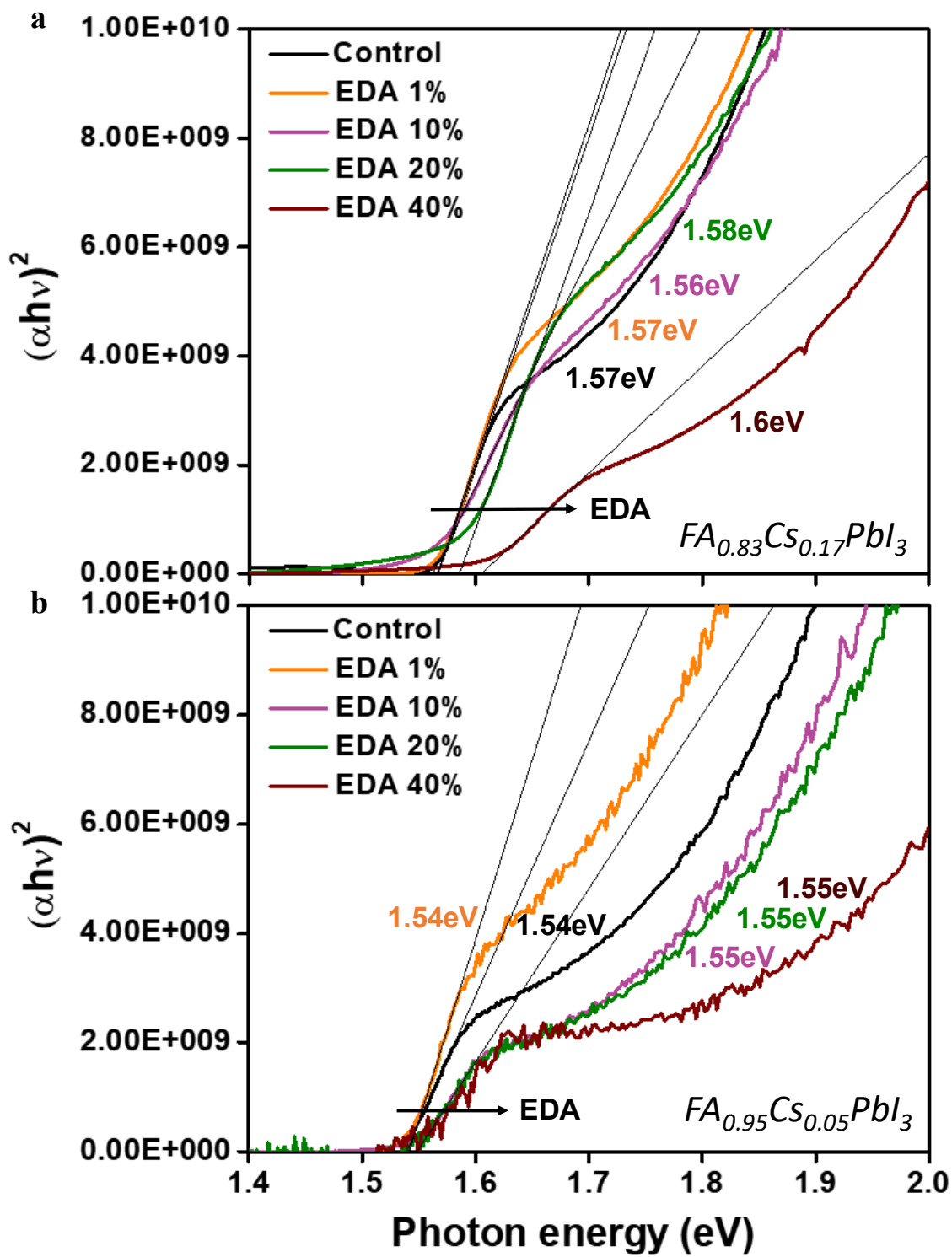

*Figure S20*: *(a) Tauc plot of FA0.83Cs0.17PbI3 and (b) FA0.95Cs0.05PbI3 with EDA 0-40 mol% addition*

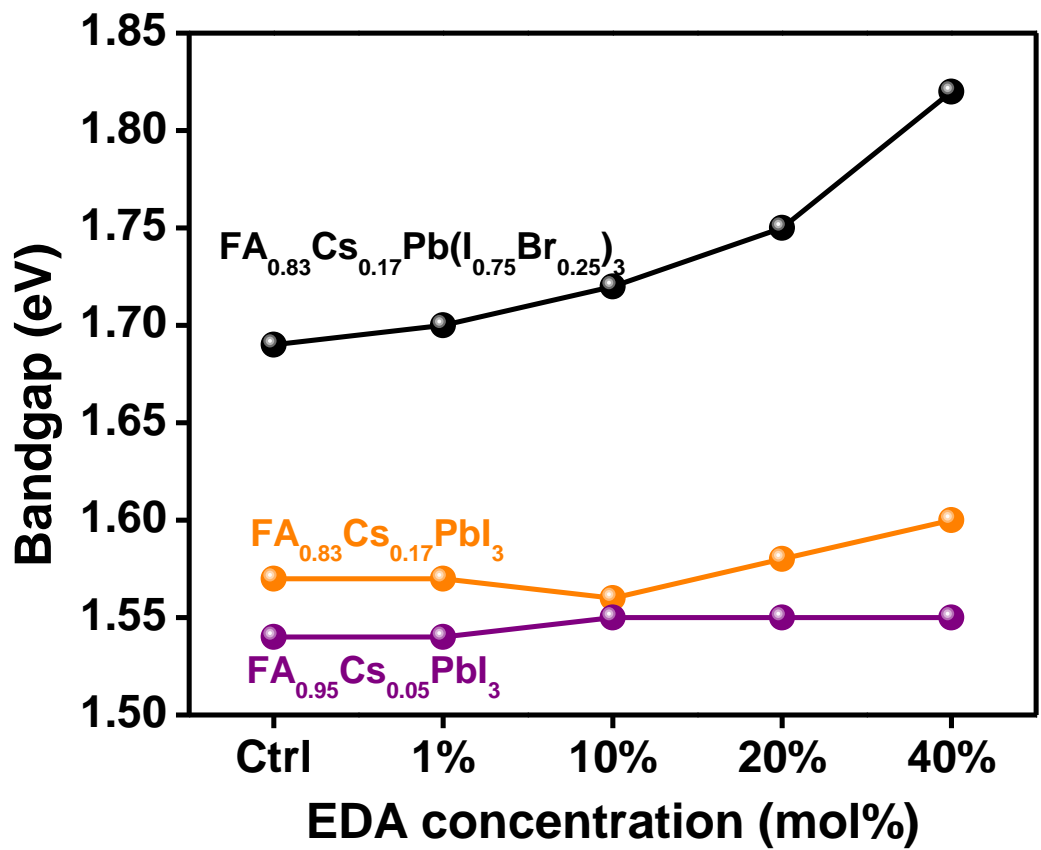

*Figure S21*: Bandgap extracted from Tauc plot shown in FigureS17 (black), Figure S20a (orange) and Figure S20b (purple) as a function of amount of EDA added in the precursor solution in mol%.

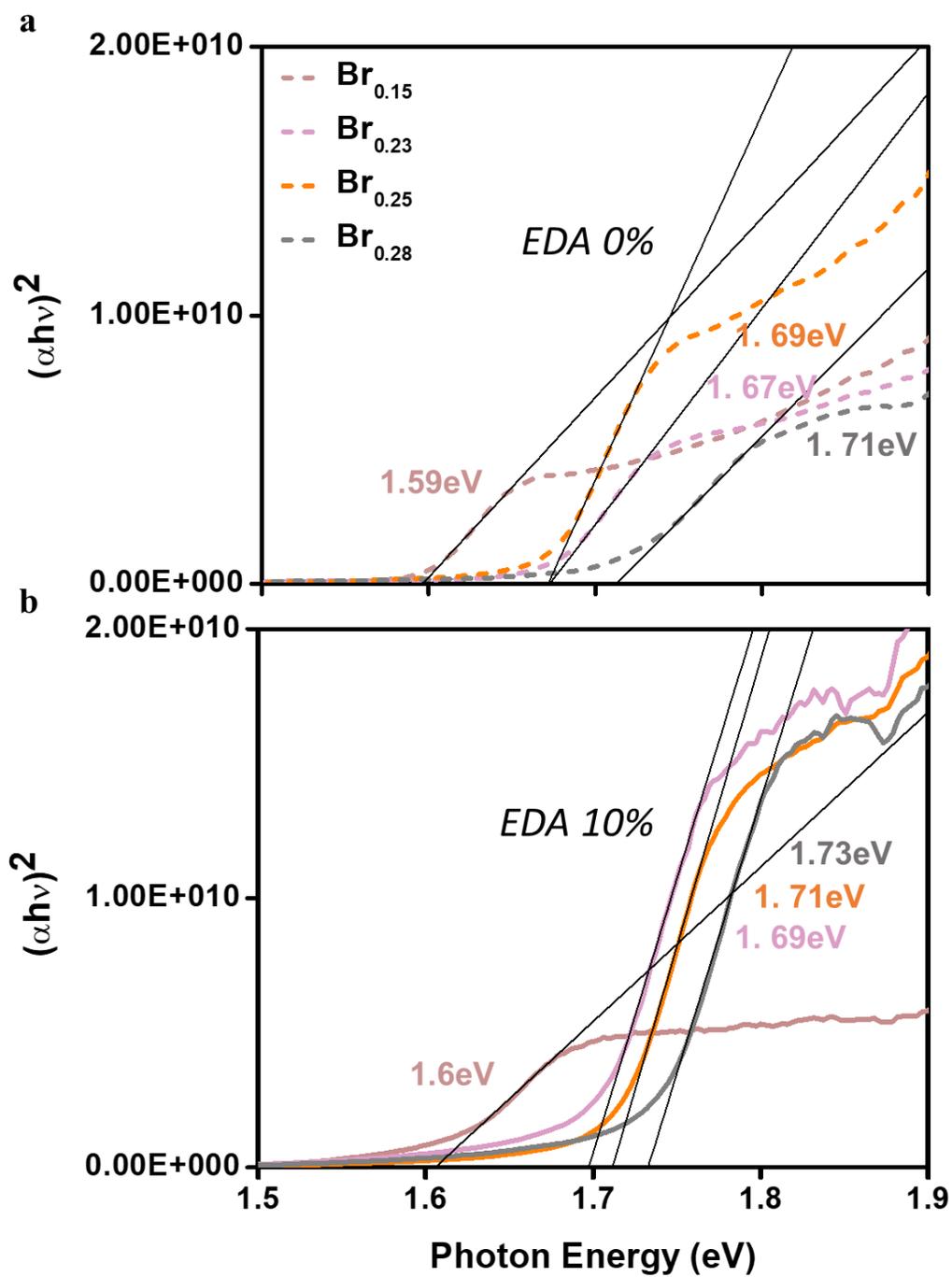

*Figure S22*: *(a) Tauc plots of the EDA-0 and (b) EDA-10 FA$_{0.83}$Cs$_{0.17}$Pb(I$_{1-x}$Br$_x$)$_3$ with x=0.15 (Br$_{0.15}$, brown), 0.23 (Br$_{0.23}$, pink), 0.25 (Br$_{0.25}$, orange), 0.28 (Br$_{0.28}$, grey)*

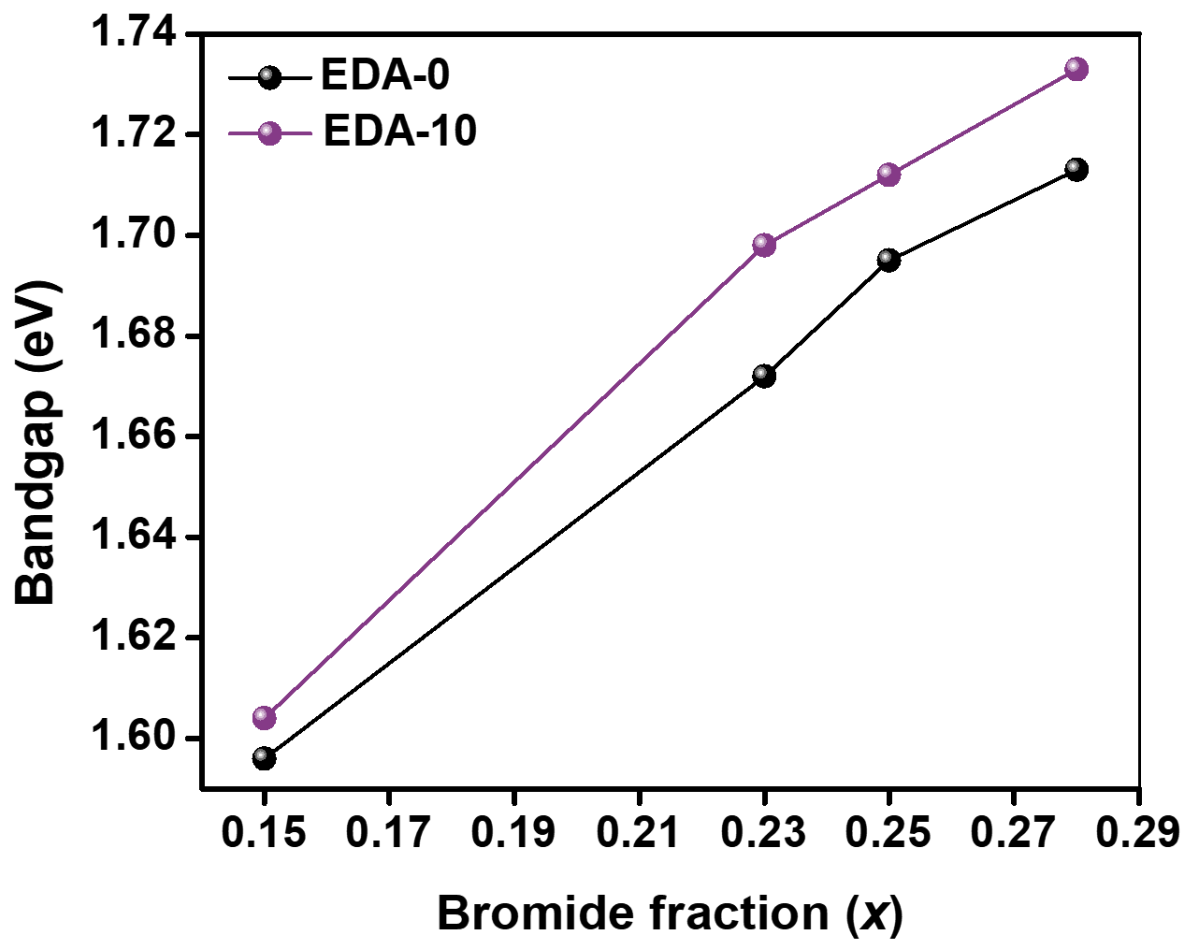

*Figure S23*: Bandgap as a function of bromide content in $FA_{0.83}Cs_{0.17}Pb(I_{1-x}Br_x)_3$ without EDA (EDA-0, black) and with EDA 10 mol% addition (EDA-10, purple) extracted from Tauc plots of FigureS22.

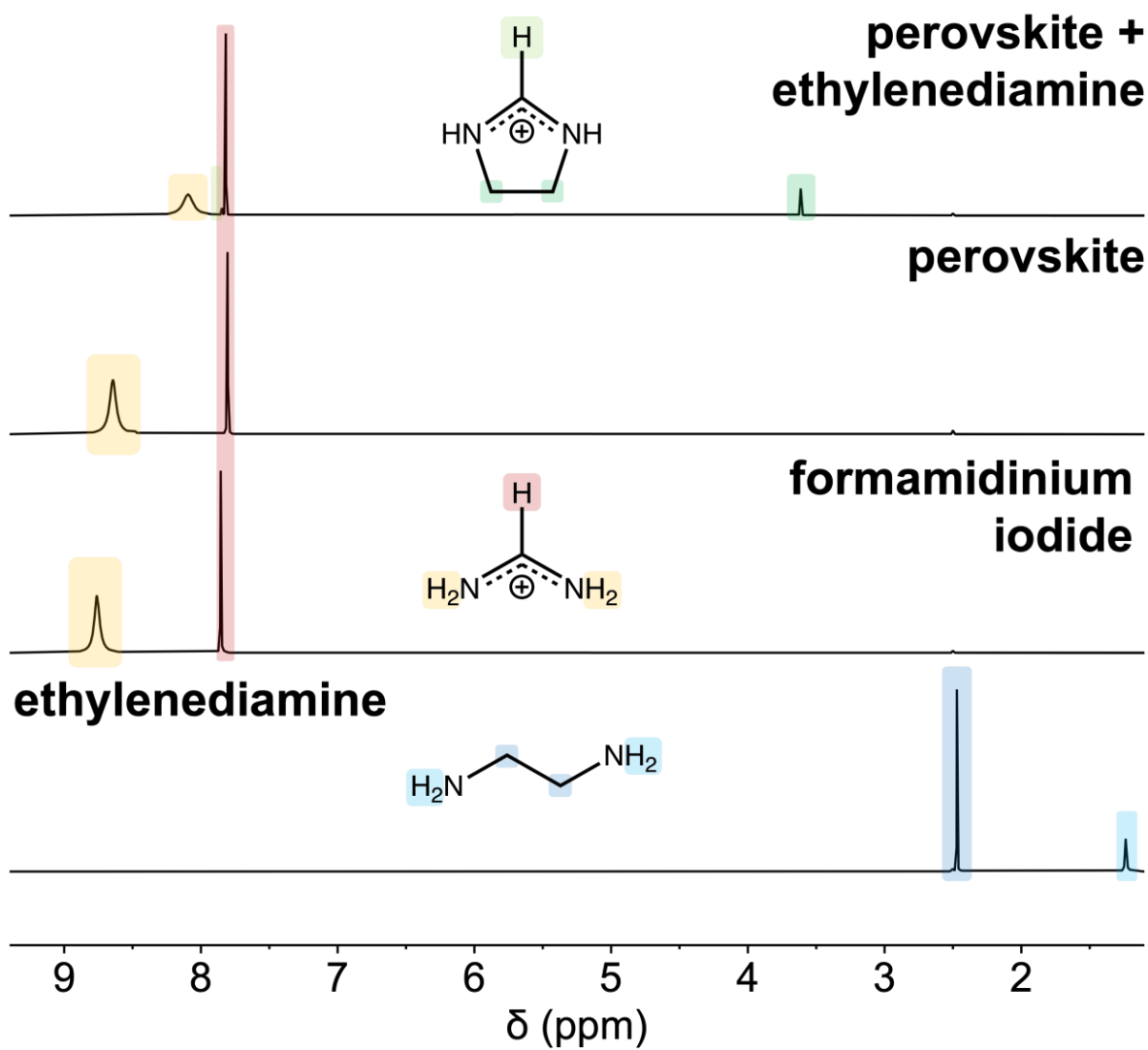

*Figure S24:* *Full spectra corresponding to those shown in main text* *Figure 4a*. *$^1$H solution NMR spectra of (bottom to top) ethylenediamine (EDA), formamidinium iodide and $(FA_{0.83}Cs_{0.17})Pb(I_{0.75}Br_{0.25})_3$ without and with 10 mol% EDA added. Signals at 2.50 ppm correspond to DMSO incidentally introduced with the deuterated solvent.*

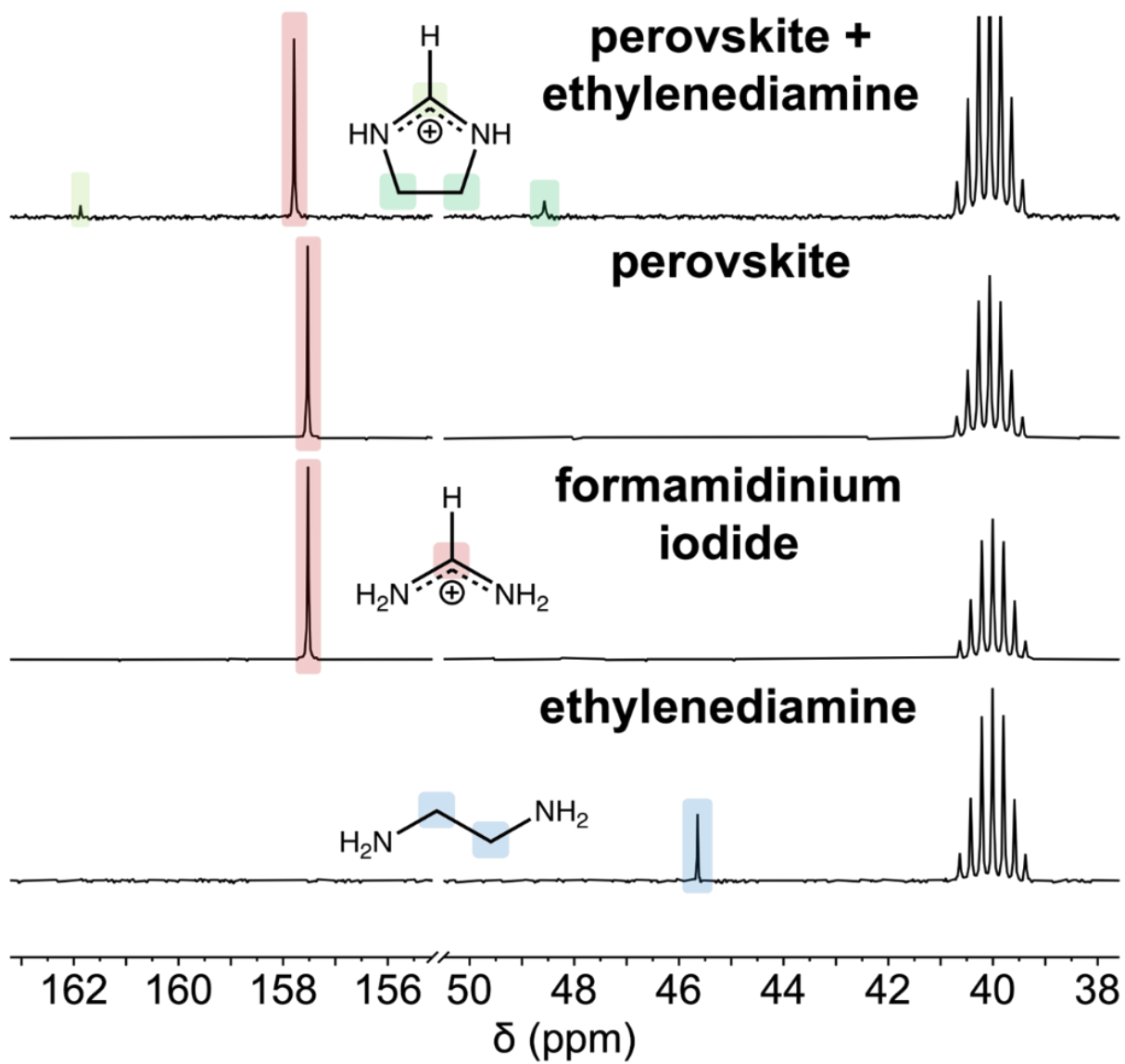

*Figure S25:* $^{13}C$ solution NMR spectra of (bottom to top) ethylenediamine (EDA), formamidinium iodide and $(FA_{0.83}Cs_{0.17})Pb(I_{0.75}Br_{0.25})_3$ without and with 10 mol% EDA added. Signals at 40.0 ppm correspond to the solvent $d^6$-DMSO.

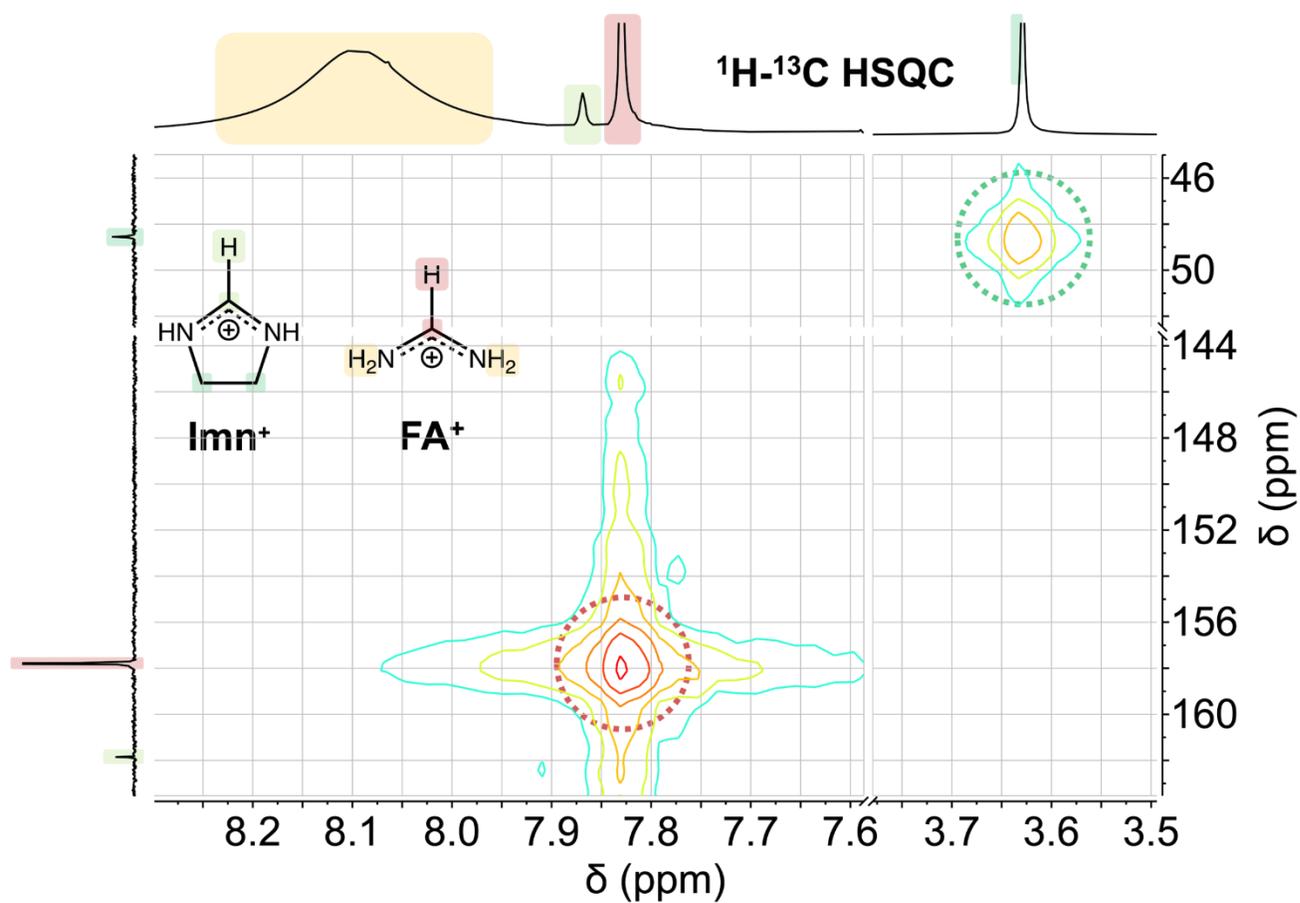

***Figure S26:*** *$^1H$-$^{13}C$ heteronuclear single quantum correlation spectroscopy (HSQC) of $(FA_{0.83}Cs_{0.17})Pb(I_{0.75}Br_{0.25})_3$ with 10 mol% EDA added. Signals in the 2D spectrum indicate that the corresponding proton and carbon environments are directly bonded.*

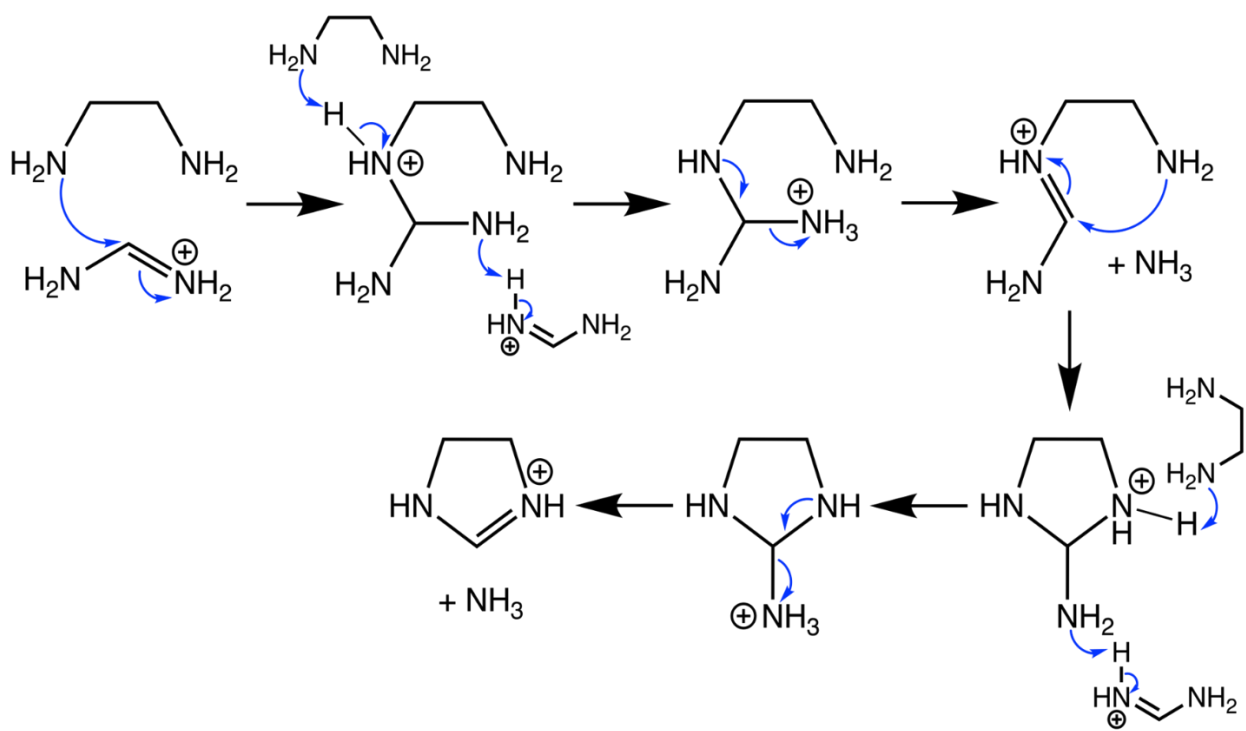

*Figure S27:* Full mechanistic justification for production of 2-imidazolinium (Imn$^+$) from ethylenediamine (EDA) and formamidinium (FA$^+$). A first nucleophilic addition of EDA to FA$^+$ is followed by reformation of the amidinium group by elimination of NH$_3$, which is highly volatile and therefore likely lost from solution to the gas phase. Rapid intramolecular nucleophilic attack by the remaining amine on the reformed amidinium system leads to cyclisation and is again followed by elimination of NH$_3$ and reformation of the conjugated amidinium group. Imn$^+$ is expected as the favoured final product of this reaction as (a) its formation is entropically favourable due to the loss of two equivalents of gaseous NH$_3$, (b) reformation of the conjugated amidinium system is thermodynamically favourable, and (c) intramolecular ring-closure is kinetically preferred to intermolecular reaction with further FA$^+$

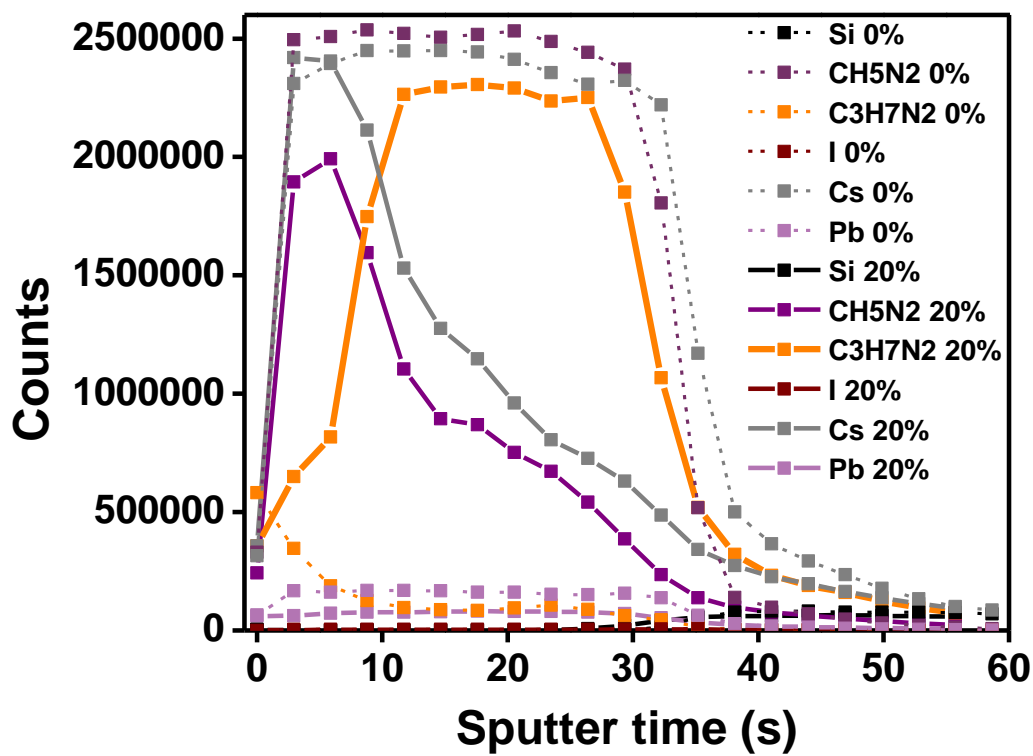

*Figure S28*: *ToF-SIMS depth profiling of control (dashed lines) and EDA-20 (solid lines). Peaks corresponding to Si (black), FA-CH5N2(purple),2-imidazolinium-C3H7N2 (orange), Iodide (dark red), cesium (grey), lead (violet) are displayed for both compositions as a function of sputter time. By looking the thickness of the film via profilometry (~480 nm) and the Si signal maximum (38 s) we can estimate the penetration depth inside the film per second of measurement (12.6 nm/s).*

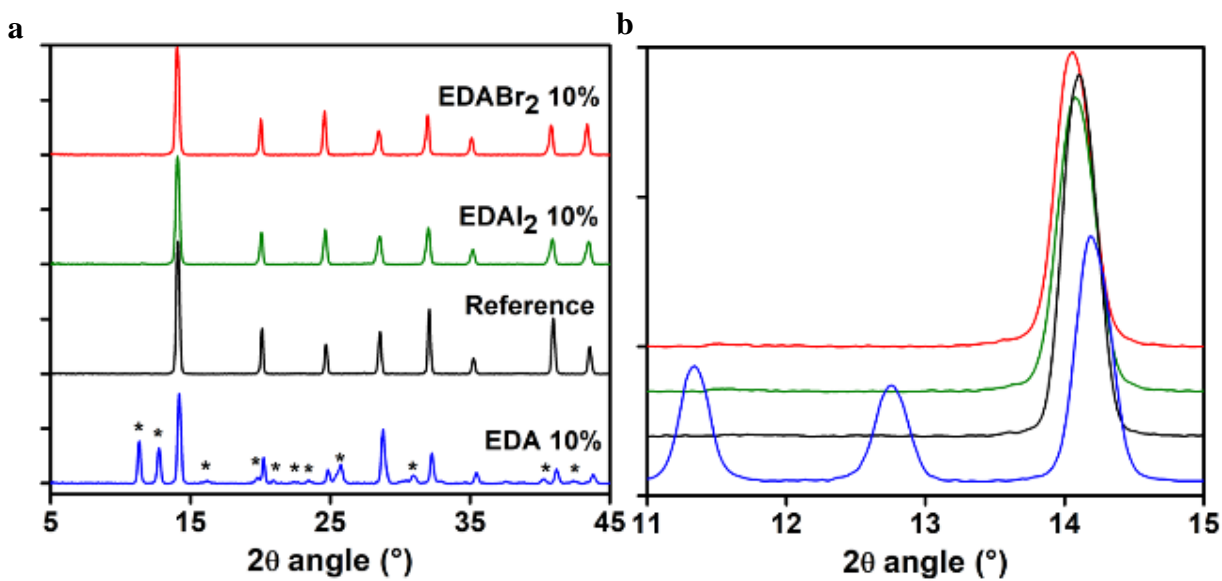

*Figure S29:* *(a) XRD patterns of the reference film (black), after addition of 10 mol% of EDAH$_2$Br$_2$ (red), EDAH$_2$I$_2$ (green) and EDA (blue). Peaks denoted with * correspond to the 4H polytype phase. (b) Highlight on the region from 11° to 15° of the patterns shown in (a).*

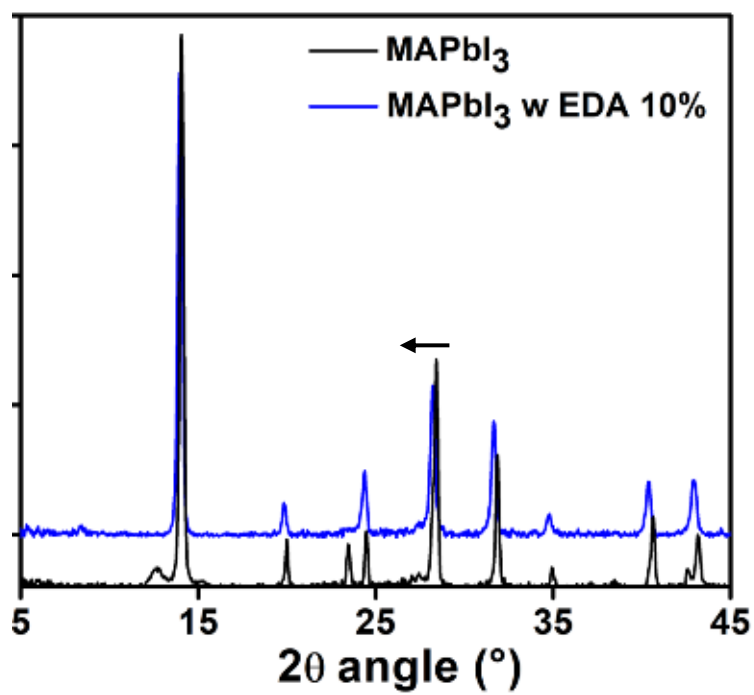

*Figure S30:* *XRD pattern of MAPbI$_3$(black) and with the addition of 10 mol% EDA (blue). The arrows show the shift to lower angles with 10% EDA.*

# References

(1) Jariwala, S.; Sun, H.; Adhyaksa, G. W. P.; Lof, A.; Muscarella, L. A.; Ehrler, B.; Garnett, E. C.; Ginger, D. S. Local Crystal Misorientation Influences Non-Radiative Recombination in Halide Perovskites. *Joule* **2019**, *3* (12), 3048–3060.

(2) deQuilettes, D. W.; Koch, S.; Burke, S.; Paranji, R. K.; Shropshire, A. J.; Ziffer, M. E.; Ginger, D. S. Photoluminescence Lifetimes Exceeding 8 Ms and Quantum Yields Exceeding 30% in Hybrid Perovskite Thin Films by Ligand Passivation. *ACS Energy Lett.* **2016**, *1* (2), 438–444.

(3) Momma, K.; Izumi, F. VESTA 3 for Three-Dimensional Visualization of Crystal, Volumetric and Morphology Data. *J. Appl. Crystallogr.* **2011**, *44* (6), 1272–1276.

(4) Al-Ashouri, A.; Köhnen, E.; Li, B.; Magomedov, A.; Hempel, H.; Caprioglio, P.; Márquez, J. A.; Morales Vilches, A. B.; Kasparavicius, E.; Smith, J. A.; Phung, N.; Menzel, D.; Grischek, M.; Kegelmann, L.; Skroblin, D.; Gollwitzer, C.; Malinauskas, T.; Jošt, M.; Matič, G.; Rech, B.; Schlatmann, R.; Topič, M.; Korte, L.; Abate, A.; Stannowski, B.; Neher, D.; Stolterfoht, M.; Unold, T.; Getautis, V.; Albrecht, S. Monolithic Perovskite/Silicon Tandem Solar Cell with >29% Efficiency by Enhanced Hole Extraction. *Science* **2020**, *370* (6522), 1300–1309.

(5) Kieffer, J.; Karkoulis, D. PyFAI, a Versatile Library for Azimuthal Regrouping. *J. Phys. Conf. Ser.* **2013**, *425* (20), 202012.

(6) Gratia, P.; Zimmermann, I.; Schouwink, P.; Yum, J.-H.; Audinot, J.-N.; Sivula, K.; Wirtz, T.; Nazeeruddin, M. K. The Many Faces of Mixed Ion Perovskites: Unraveling and Understanding the Crystallization Process. *ACS Energy Lett.* **2017**, *2* (12), 2686–2693.

(7) Zhang, Z.; Zhu, Y.; Wang, W.; Zheng, W.; Lin, R.; Huang, F. Growth, Characterization and Optoelectronic Applications of Pure-Phase Large-Area $CsPb_2Br_5$ Flake Single Crystals. *J. Mater. Chem. C* **2018**, *6* (3), 446–451.

(8) Hu, Y.; Aygüler, M. F.; Petrus, M. L.; Bein, T.; Docampo, P. Impact of Rubidium and Cesium Cations on the Moisture Stability of Multiple-Cation Mixed-Halide Perovskites. *ACS Energy Lett.* **2017**, *2* (10), 2212–2218.

(9) Glushkova, A.; Arakcheeva, A.; Pattison, P.; Kollár, M.; Andričević, P.; Náfrádi, B.; Forró, L.; Horváth, E. Influence of the Organic Cation Disorder on Photoconductivity in Ethylenediammonium Lead Iodide, $NH_3CH_2CH_2NH_3PbI_4$. *CrystEngComm* **2018**, *20* (25), 3543–3549.

(10) Spanopoulos, I.; Ke, W.; Stoumpos, C. C.; Schueller, E. C.; Kontsevoi, O. Y.; Seshadri, R.; Kanatzidis, M. G. Unraveling the Chemical Nature of the 3D "Hollow" Hybrid Halide Perovskites. *J. Am. Chem. Soc.* **2018**, *140* (17), 5728–5742.

(11) Haris, M. P. U.; Bakthavatsalam, R.; Shaikh, S.; Kore, B. P.; Moghe, D.; Gonnade, R. G.; Sarma, D. D.; Kabra, D.; Kundu, J. Synthetic Control on Structure/Dimensionality and

Photophysical Properties of Low Dimensional Organic Lead Bromide Perovskite. *Inorg. Chem.* **2018**, *57* (21), 13443–13452.